\documentclass{99-Styles/MICE}
\usepackage{bm} 
\usepackage{amsmath}
\usepackage{graphicx}

\usepackage{float}
\floatstyle{plaintop}
\restylefloat{table}
\usepackage[tableposition=top]{caption}

\usepackage{lineno}
\usepackage{bibentry}
\nobibliography*
\usepackage{times}
\usepackage{bm}         
\usepackage{amsmath,amssymb,bm,slashed} 
\usepackage{amssymb}    
\usepackage{graphicx}   
\usepackage{verbatim}   
\usepackage{color}      
\usepackage{subfigure}  
\usepackage{hyperref}   
\usepackage{enumerate}

\usepackage{fmtcount}   

\usepackage[square,comma,numbers,sort&compress]{natbib}

\begin{document}

\thispagestyle{empty}

\begin{tabular}{p{0.175\textwidth} p{0.5\textwidth} p{0.225\textwidth}}
  \hspace{-0.8cm}\leftline{\today}                                 &
  \centering{Muon Ionization Cooling Experiment}                  &
  \rightline{}
\end{tabular}
\vspace{-1.0cm}\\
\rule{\textwidth}{0.43pt}

\begin{center}
  {\bf
    {\LARGE The reconstruction software for the MICE scintillating fibre trackers } \\
  }
  \vspace{0.2cm}
    A.~Dobbs, C.~Hunt, K.~Long, E.~Santos\footnote{Now at Winton Capital Management}, M.~A.~Uchida
  \\{\it
    Physics Department, Blackett Laboratory, Imperial College London \\
    Exhibition Road, London, SW7 2AZ, UK
  }
  \par 
  P.~Kyberd
  \\{\it
    Brunel University London, Kingston Lane, Uxbridge, Middlesex, UB8 3PH, U.K.
  } \\ 
  \par 
  C.~Heidt
  \\{\it
    University of California Riverside, 900 University Ave, Riverside, CA 92521, U.S.A. 
  }\\
  \par
  S.~Blot\footnote{Now at Deutsches Elektronen-Synchrotron}
  \\{\it
    University of Chicago, Edward H. Levi Hall, 5801 South Ellis Avenue, Chicago, Illinois, U.S.A. 
  }\\
  \par
  E.~Overton
  \\{\it
    University of Sheffield, Western Bank, Sheffield, S10 2TN, U.K. 
  }\\
  \par 
\end{center}


\newcommand{\bra}[1]{\ensuremath{\langle #1 |}}   
\newcommand{\ket}[1]{\ensuremath{| #1 \rangle}}   
\newcommand{\bigbra}[1]{\ensuremath{\big\langle #1 \big|}}   
\newcommand{\bigket}[1]{\ensuremath{\big| #1 \big\rangle}}   
\newcommand{\amp}[3]{\ensuremath{\left\langle #1 \,\left|\, #2%
                     \,\right|\, #3 \right\rangle}}  
\newcommand{\sprod}[2]{\ensuremath{\left\langle #1 |%
                     #2 \right\rangle}}  
\newcommand{\ev}[1]{\ensuremath{\left\langle #1 %
                     \right\rangle}} 
\newcommand{\ds}[1]{\ensuremath{\! \frac{d^3#1}{(2\pi)^3 %
                     \sqrt{2 E_\vec{#1}}} \,}} 
\newcommand{\dst}[1]{\ensuremath{\! %
                     \frac{d^4#1}{(2\pi)^4} \,}} 
\newcommand{\tr}{\text{tr}}
\newcommand{\sgn}{\text{sgn}}
\newcommand{\diag}{\text{diag}}
\newcommand{\BR}{\text{BR}}

\renewcommand{\vec}[1]{{\mathbf{#1}}}
\renewcommand{\Re}{{\text{Re}}}
\renewcommand{\Im}{{\text{Im}}}
\newcommand{\iso}[2]{{\ensuremath{{}^{#2}}\ensuremath{\rm #1}}}
\newcommand{\eps}{{\ensuremath{\epsilon}}}
\newcommand{\draftnote}[1]{{\bf\color{red} \MakeUppercase{#1}}}
\newcommand{\panm}[1]{{\color{blue} #1}}
\providecommand{\abs}[1]{\lvert#1\rvert}
\providecommand{\norm}[1]{\lVert#1\rVert}

\def\parenbar{\mathpalette\p@renb@r}
\def\p@renb@r#1#2{\vbox{%
  \ifx#1\scriptscriptstyle \dimen@.7em\dimen@ii.2em\else
  \ifx#1\scriptstyle \dimen@.8em\dimen@ii.25em\else
  \dimen@1em\dimen@ii.4em\fi\fi \offinterlineskip
  \ialign{\hfill##\hfill\cr
    \vbox{\hrule width\dimen@ii}\cr
    \noalign{\vskip-.3ex}%
    \hbox to\dimen@{$\mathchar300\hfil\mathchar301$}\cr
    \noalign{\vskip-.3ex}%
    $#1#2$\cr}}}

%
\providecommand{\anmne}{\mbox{$\bar\nu_{\mu} \rightarrow \bar\nu_e$}} 
\providecommand{\nmne}{\mbox{$\nu_{\mu}\rightarrow\nu_e$}} 
\providecommand{\anm}{\mbox{$\bar\nu_\mu$}} 
\providecommand{\nm}{\mbox{$\nu_\mu$}}
\providecommand{\nue}{\mbox{$\nu_e$}} 
\providecommand{\ane}{\mbox{$\bar\nu_e$}} 
\providecommand{\enu}{\mbox{$E_\nu$}}
\providecommand{\piz}{\mbox{$\pi^0 $}}
\providecommand{\pip}{\mbox{$\pi^+$}} 
\providecommand{\pim}{\mbox{$\pi^-$}}

\parindent 10pt
\pagestyle{plain}
\pagenumbering{arabic}                   
\setcounter{page}{1}

\begin{quotation}

\noindent
The Muon Ionization Cooling Experiment (MICE) will demonstrate the principle of muon beam phase-space reduction via ionization cooling.  Muon beam cooling will be required for the proposed Neutrino Factory or Muon Collider.  The phase-space before and after the cooling cell must be measured precisely. This is achieved using two scintillating-fibre trackers, each placed in a solenoidal magnetic field.  This paper describes the software reconstruction for the fibre trackers: the GEANT4 based simulation; the implementation of the geometry; digitisation; space-point reconstruction; pattern recognition; and the final track fit based on a Kalman filter. The performance of the software is evaluated by means of Monte Carlo studies and the precision of the final track reconstruction is evaluated.

\end{quotation}

\section{The MICE Experiment}
\label{sec:MICE}
  \subsection{Overview}
  \label{subsec:Overview}
  The Muon Ionization Cooling Experiment (MICE) will perform a practical demonstration of muon ionization cooling. Cooling refers to a reduction in the emittance of a beam, that is, the reduction of the phase-space volume occupied by the beam. Beam cooling is required for any future facility based on high intensity muon beams, such as a Neutrino Factory~\cite{ISS-Physics}, the ultimate tool to study leptonic CP-invariance violation, or a Muon Collider~\cite{MC_Overview}, a potential route to multi-TeV lepton -- anti-lepton collisions. Muon beams are generated via pion decay, and therefore have a large emittance, which must be reduced so that a reasonable fraction of the beam will fall within the acceptance of the downstream acceleration system.

  The short muon lifetime requires fast beam cooling that traditional techniques are unable to provide.  Ionization cooling was proposed in the early 1970s~\cite{Skrinsky, Neuffer}, but has not yet been demonstrated at the energies of interest for the Neutrino Factory or Muon Collider.  Ionization cooling reduces emittance by passing a beam through some suitable material of low atomic-number such as liquid hydrogen.  This leads to the reduction of all components of momentum due to ionization energy loss. Low atomic number absorbers are preferred because they minimise multiple scattering which ``heats'' the beam. 

  \begin{figure}[bht]
    \begin{center}
      \includegraphics[width=1.0\linewidth]{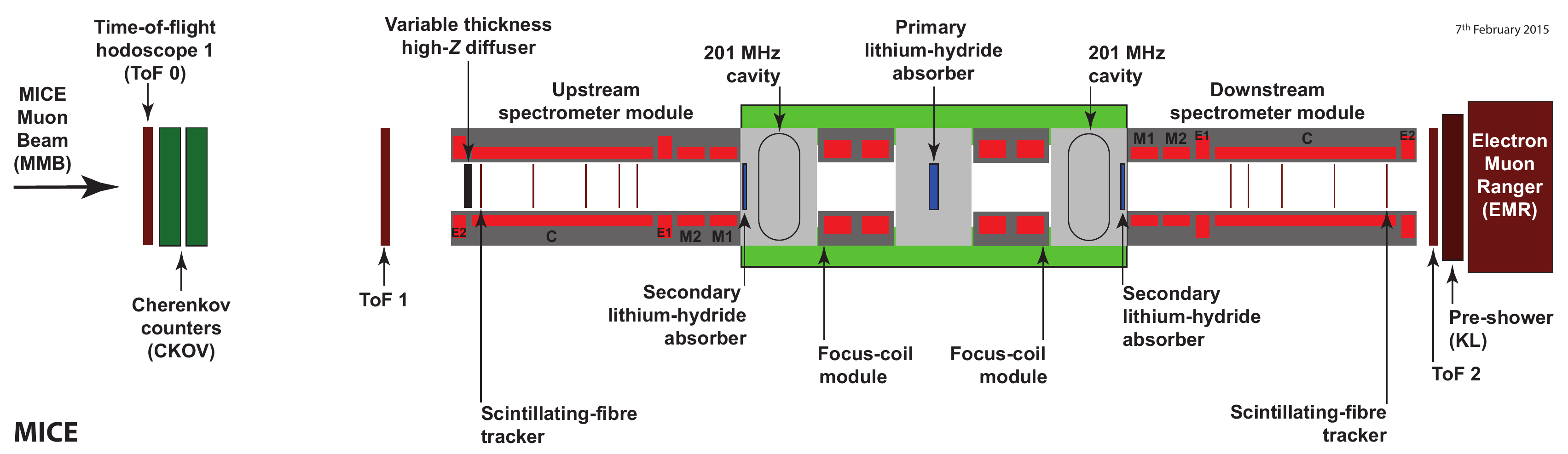}
      \caption{\label{fig:CoolingChannel} The beam diagnostics and cooling cell (indicated in green). $M$ refers to matching coils, $E$ to end coils and $C$ to the centre coils. Particle identification is provided by three time-of-flight stations, two threshold Cherenkov detectors and a downstream calorimeter composed of a pre-shower detector and an electron-muon ranger. Emittance is measured upstream and downstream of the cooling cell using spectrometers. A diffuser is used to increase the beam emittance prior to the upstream spectrometer. Ionization energy loss occurs as the beam passes through the absorbers, while acceleration is providing by radio frequency cavities.}
    \end{center}
  \end{figure}

  MICE is sited at the Science and Technology Facilities Council Rutherford Appleton Laboratory in the U.K. and exploits the ISIS proton synchrotron to generate the muon beam~\cite{MiceTarget}.  The MICE beam line is described in detail in~\cite{BeamlineJINST}. A schematic of the full MICE experiment is shown in figure~\ref{fig:CoolingChannel}. The beamline is instrumented with with two threshold Cherenkov detectors, three time-of-flight (TOF) stations and a downstream calorimeter. The cooling cell consists of three absorber modules and two radio frequency (RF) cavities. Two high-precision trackers in a solenoidal field are used to measure the emittance change (see section~\ref{subsec:Trackers}).

  MICE is a staged experiment, built and run in distinct steps. The first step of the programme, consisting of the muon beam line with particle identification, is now complete and results are available in~\cite{BeamlineJINST, BeamCharacterisationEurPhysJ, EMRJINST, PionContaminationJINST}. The present step of the program, which introduced the trackers and the first absorber module, began taking data in 2015. 


  \subsection{The Scintillating Fibre Trackers}
  \label{subsec:Trackers}
\begin{figure}[tb]
    \centering
    \includegraphics[width=0.5\linewidth]{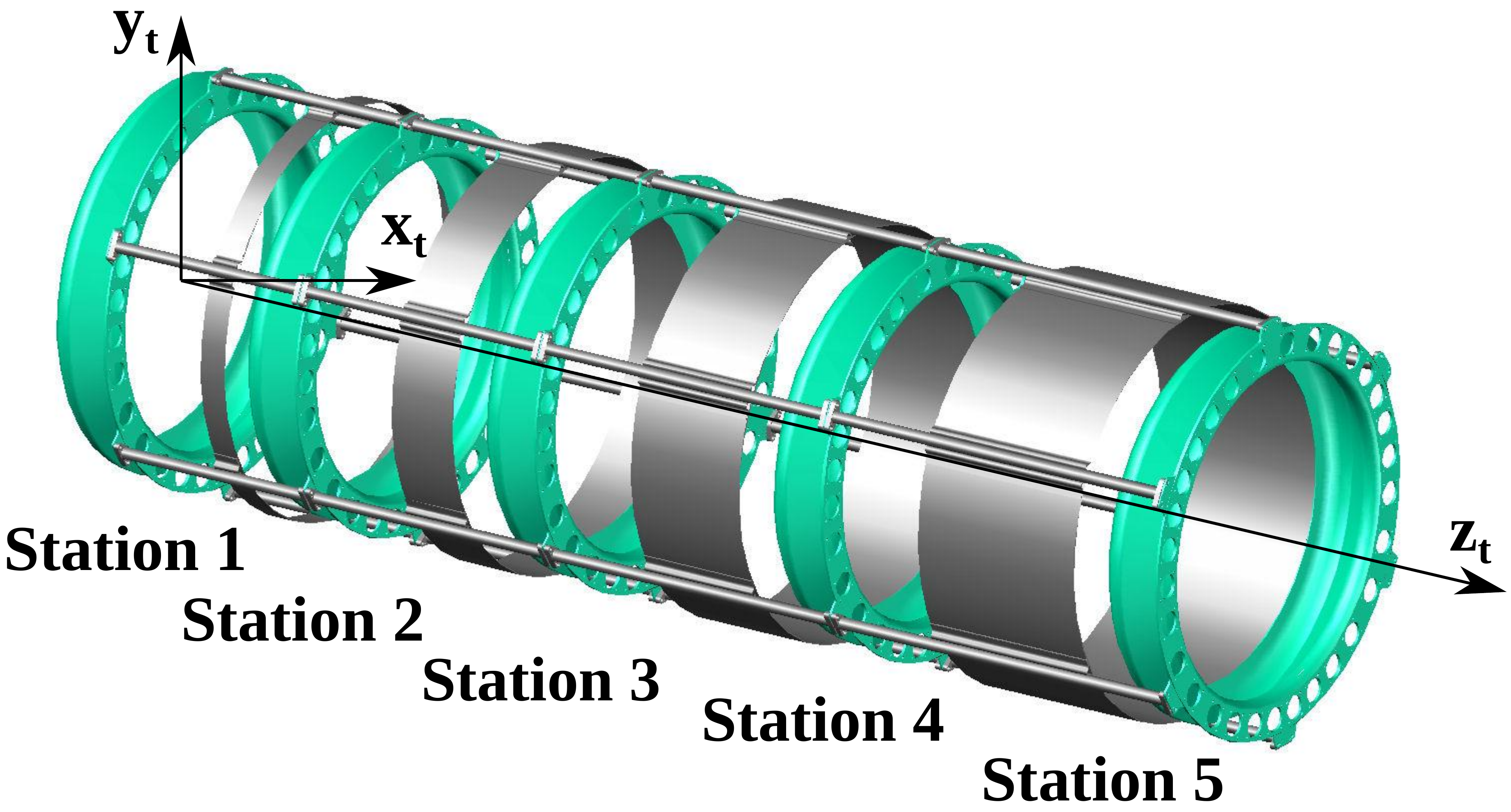} \hspace{2pc}%
    \includegraphics[width=0.35\linewidth]{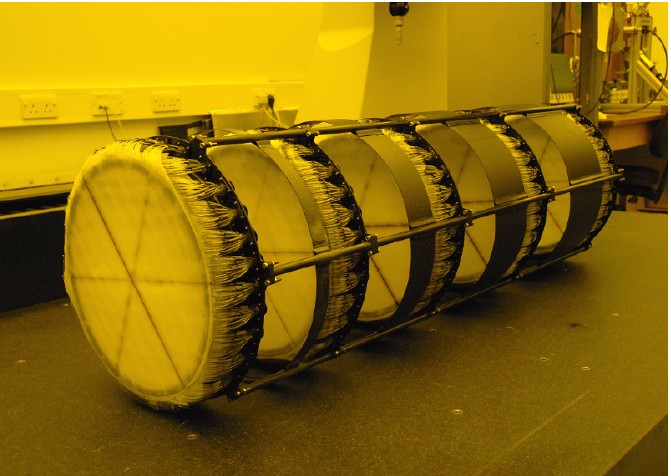}
    \caption{\label{fig:Trackers} Left: A schematic of the tracker carbon-fibre frame.  The fibre planes are glued on to the upstream edge of the carbon-fibre station frames (shown in green). The longitudinal coordinate of the tracker coordinate system ($z_t$) increases as one moves from the fibre plane towards the station frame to which it is attached.  Right: A photograph of a tracker. The colour is due to the filtered lighting needed to protect the scintillating-fibres. The intersecting lines visible on the station faces indicate the direction of the fibres in each plane.}
  \end{figure}

  MICE is equipped with two identical, high precision scintillating-fibre (``scifi'') trackers, described in~\cite{TrackersNIM}. Each tracker is placed in a superconducting solenoid that provides a uniform field over the tracking volume. One tracker, TKU, is upstream of the cooling cell, the other, TKD, downstream.  Each tracker consists of 5 detector stations, labelled 1 to 5, as illustrated in figure~\ref{fig:Trackers}. TKU is orientated such that Station~5 sees the beam first, TKD is rotated by 180$^\circ$ such that Station~1 sees the beam first, thus, in both trackers, Station~1 is always nearest to the cooling cell (see figure~\ref{fig:CoolingChannel}).

\begin{figure}[tb]
    \begin{center}
      \includegraphics[width=0.85\textwidth]{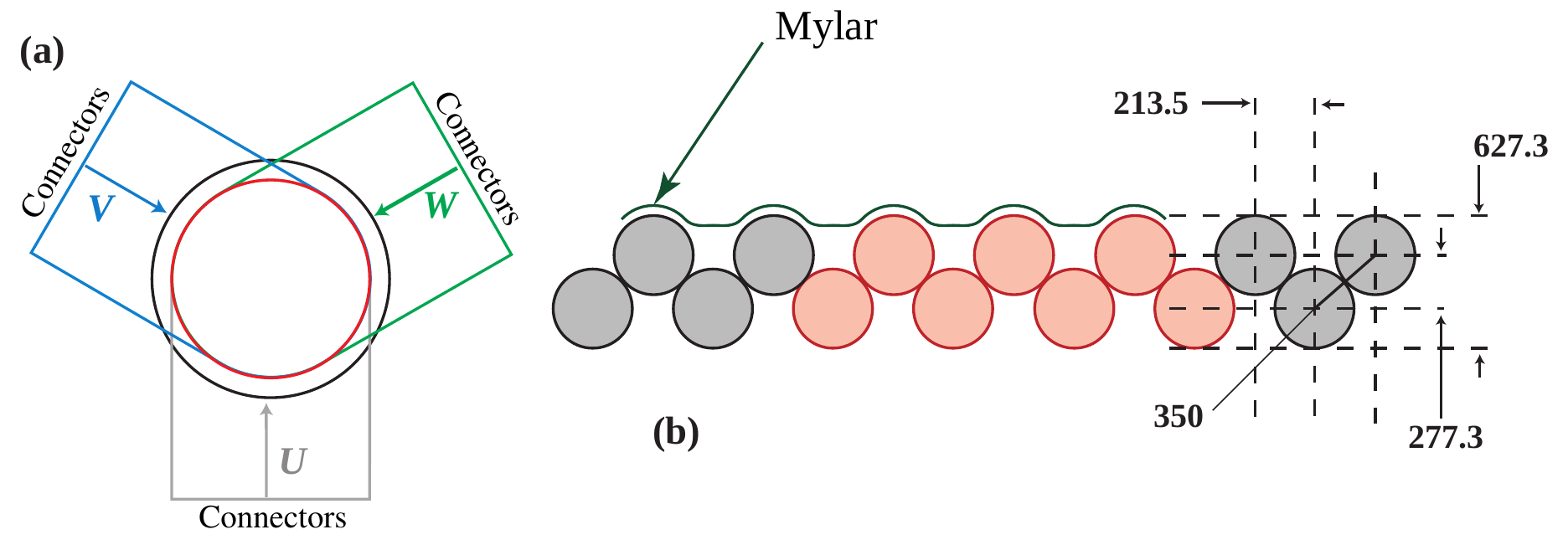}
      \caption{\label{fig:DoubletLayer}(a) Arrangement of the doublet layers in the scintillating-fibre  stations. The outer circle shows the solenoid bore while the inner circle shows the limit of the active area of the tracker. The arrows indicate the direction that the individual 350\,$\mu$m fibres run. (b) Detail of the arrangement of the scintillating-fibres in a doublet layer. The fibre spacing and the fibre pitch are indicated on the right-hand end of the figure in \,$\mu$m. The pattern of seven fibres shown in red form a single channel, which is readout via a clear-fibre light-guide. The sheet of mylar glued to the doublet layer is indicated. }
    \end{center}
  \end{figure}

\begin{figure}[tb]
    \centering
    \includegraphics[width=0.95\linewidth]{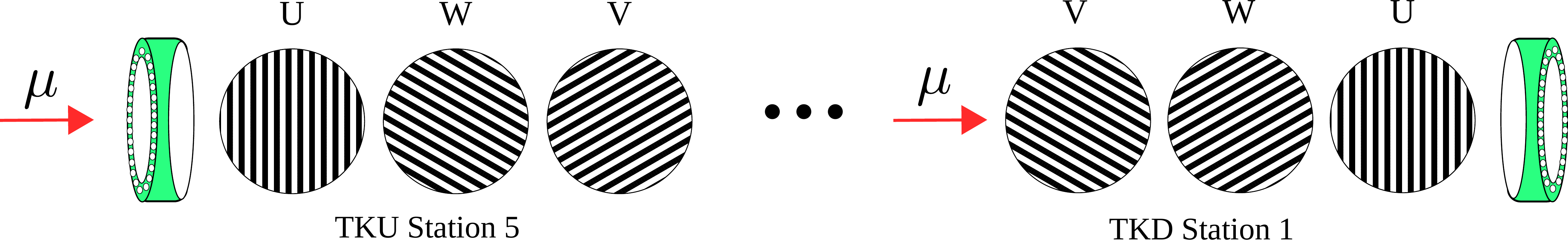} \hspace{2pc}%
    \caption{\label{fig:FibrePlaneOrientation} The orientation of the fibres in each plane, as seen by the incoming beam, for both trackers. The green object is the station frame.}
  \end{figure}

  Each station is formed of three planes of 350~$\mu$m scintillating-fibres, orientated at 120 degrees to one another. The fibres in each plane are arranged in two layers offset with respect to each other (a ``doublet layer''), in order to give 100$\%$ coverage of the plane area as illustrated in figure~\ref{fig:DoubletLayer}. The doublet layer is glued on to a sheet of mylar. The fibres are collected into groups of seven for readout, each group forming a single channel, as illustrated in figure~\ref{fig:DoubletLayer}b. The planes, also known as views, are labelled $U$, $V$ and $W$. Plane $U$ is attached to the station frame directly, plane $W$ on to plane $U$, and plane $V$ on to plane $W$. The fibre-plane orientations are illustrated in figure~\ref{fig:FibrePlaneOrientation}. Each station is oriented such that the fibres in the $U$ plane are vertical. The fibres produce scintillation light when ionizing radiation passes through them. Clear-fibre light guides transport the scintillation light to visible light photon counters (VLPCs) that are operated at 9~$K$ in a cryostat~\cite{TrackersNIM}. The signal from the VLPCs is digitised using front-end electronics developed by the D0 experiment~\cite{D0}. 
  
\section{Coordinate systems and reference surfaces}
\label{sec:Coordinates}

  The planes, stations and trackers themselves each have a coordinate system defined as described below. Coordinates with subscript $p$ will refer to the plane coordinate system, $s$ to the station coordinate system and $t$ to the tracker coordinate system.

  \subsection{Channels and digits}
\begin{figure}[htb]
    \begin{center}
      \includegraphics[width=0.9\textwidth]{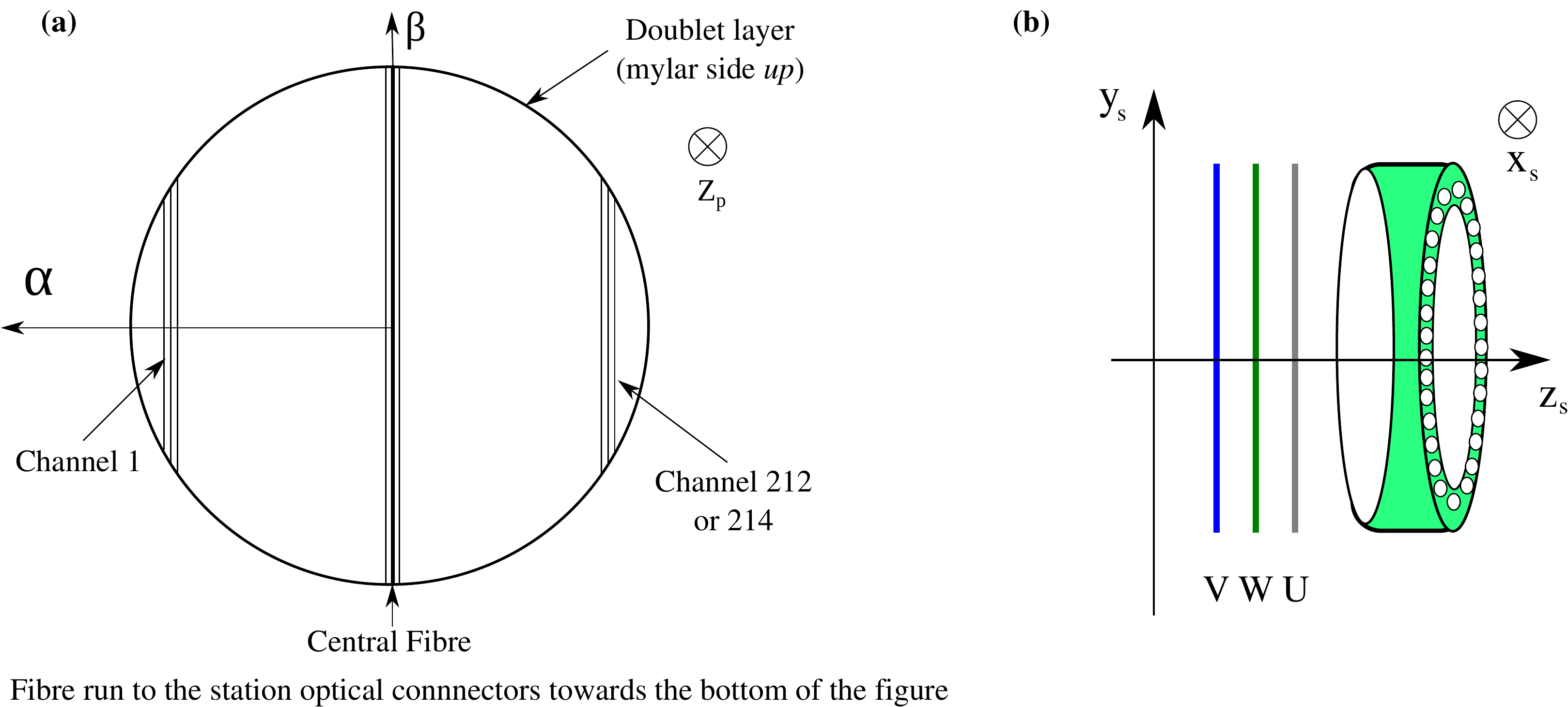}
      \caption{\label{fig:DoubletLayerOrder} (a)~The channel numbering within a plane, and the $(\alpha, \beta, z_p)$ plane coordinate system (a right-handed system).  (b)~The fibre plane ordering with respect to the station body and the $(x_s, y_s, z_s)$ station coordinate frame (a right-handed system).  In TKU the beam approaches from the right, in TKD from the left. Note that $z_s$ is by definition equivalent to $z_p$ of the $V$ plane.}
    \end{center}
  \end{figure}

  The $V$ and $W$ planes each consist of 214 channels, labelled 0 to 213, while the $U$ plane has 212 channels, labelled 0 to 211.  The channel number increases from left to right if a plane is placed mylar side up, with the fibre readout pointing downwards, as illustrated in figure~\ref{fig:DoubletLayerOrder}.

  \subsection{Planes and clusters}
  \label{subsec:PlaneAndClusters}
  Each plane is assigned an integer known as the plane number: plane~$V$ is assigned~0; plane~$W$ is assigned~1; and plane~$U$ is assigned~2.  The plane reference surface is defined to be the flat plane that is formed by the outer surface of the mylar sheet. The measured position perpendicular to the direction of the fibres in each plane is labelled  $\alpha \in (v, u, w)$, defined to increase in the \textit{opposite} direction to the channel number with zero as the mid-point of the central channel. $\alpha$ is then given by $\alpha = \left(N_{CC} - N_{Ch}\right) \times d$, where $N_{CC}$ is the central channel number, $N_{Ch}$ is the channel number and $d$ is the channel width.
  
%
  
  The $z$ axis of the plane coordinate system is defined to be perpendicular to the plane reference surface and points in the direction from the mylar sheet towards the fibres. The direction in which the fibres run defines the final plane coordinate, $\beta$, completing a right-handed coordinate system. The origin of the $(\alpha, \beta)$ coordinate system is taken to be at the centre of the circular active area of the plane.

  \subsection{Stations and spacepoints}
  The station reference surface is defined to coincide with the reference surface of the $V$ doublet-layer. The station coordinate system is defined such that the $x_s$ axis is coincident with the $u$ coordinate (that is, $\alpha$ for the $U$ layer), the $z_s$ axis is coincident with the $z_p$ axis of the $V$ layer and the $y_s$ axis completes a right-handed coordinate system.

  \subsection{Trackers and tracks}
  Each tracker is assigned an integer known as the tracker number: TKU is assigned~0 and TKD is assigned~1. The tracker reference surface is defined to coincide with the reference surface of Station~1. The tracker coordinate system is defined such that the $z_t$ axis coincides with the axis of cylindrical symmetry of the tracker as shown in figure~\ref{fig:Trackers}. The tracker $z_t$ coordinate increases from Station~1 to Station~5. The tracker $x_t$ axis is defined to coincide with the $x_s$ axis of Station~1 and the tracker $y_t$ axis completes a right-handed coordinate system. 
  
\section{The MAUS framework}
\label{sec:MAUS}
The tracker software is part of the MICE software framework, known as MAUS (MICE Analysis User Software)~\cite{MausIPAC11}. MAUS is used to perform Monte Carlo simulation and both online and offline data reconstruction. It is built using a combination of C++ and Python, with C++ being used for more processor-intensive tasks and Python being used more in the code presented to the user.  Simulation is based on GEANT4~\cite{GEANT4}, with analysis based on ROOT~\cite{ROOT}.  ROOT files are used as the primary output data format. 

MAUS programmes are defined in a Python script together with a configuration file.  This script allows the user to create programmes by combining different MAUS modules depending on the task at hand, following the Map-Reduce programming model~\cite{MapReduce}. The object passed between the modules is known as a ``spill'', representing the data associated with one spill of particles passing through the MICE beamline (see~\cite{BeamlineJINST}).  The modules come in four types: Input; Output; Map; and Reduce.  Input modules provide the initial data to MAUS, from a data file, or from the DAQ. Maps perform most of the simulation and analysis work and may be processed in parallel across multiple nodes.  Reducers are used to display output, such as for online reconstruction plots, and are capable of accumulating data sent from maps over multiple spills, but must be run in a single thread. Output modules provide data persistency.

The tracker software consists of 7 maps and a reducer. The maps cover: digitisation of Monte Carlo data; digitisation of real DAQ data; the addition of noise to Monte Carlo data; cluster reconstruction; spacepoint reconstruction; pattern recognition; and the final track fit. The reducer provides real time information on the tracker performance.  
\section{Data structure}
\label{sec:DataStructure}

\subsection{General MAUS, Monte Carlo and DAQ data structures}
\label{subsec:GeneralDataStructure}
\begin{figure}[bt]
  \begin{center}
    \includegraphics[width=27pc]{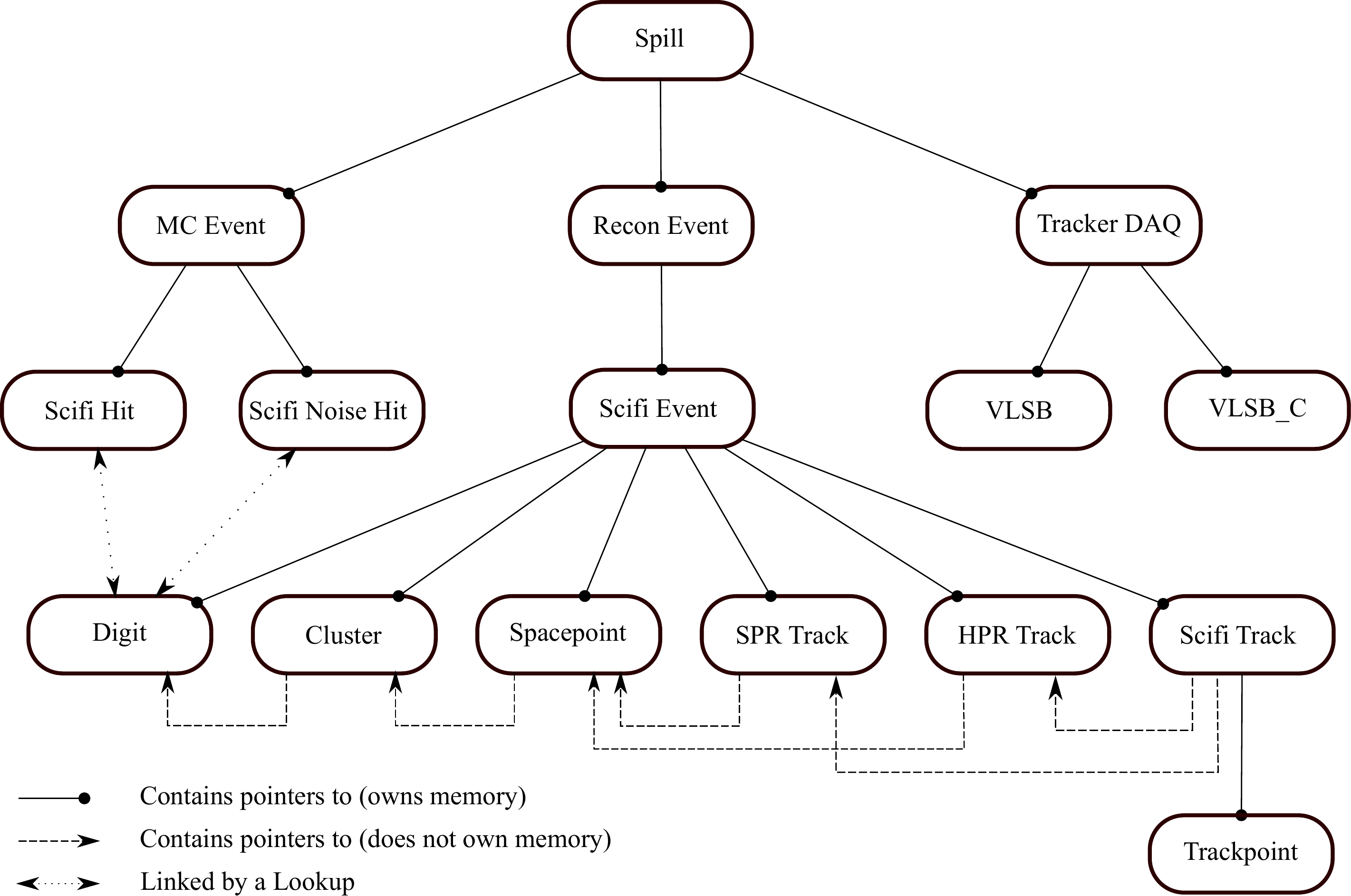}
    \caption{\label{figureDataStructure}The tracker software data structure and relevant MAUS data structure.  The spill is the top level object below which branches hold real-data, MC data and reconstructed data objects. When an object owns the memory of a set of other objects, these are held as standard vectors of pointers. When an object contains cross links to another set of objects, without owning their memory, these are held as a ROOT TRefArray of pointers. MC = Monte Carlo, SPR = Straight Pattern Recognition, HPR = Helical Pattern Recognition.}
  \end{center}
\end{figure}

A simplified schematic of the tracker data structure, with the relevant entries from the more general MAUS data structure, is shown in figure~\ref{figureDataStructure}.  All the objects listed represent container classes for different parts of the simulation, raw data and reconstruction.  The top-level object is the spill (see section~\ref{sec:MAUS}).  Within the spill the data is split into three branches: real-data from the DAQ; Monte Carlo data generated by simulation; and reconstructed data, which is formed from data in either the real or Monte Carlo branches. 

The reconstruction code makes no direct reference to the Monte Carlo information and has no way to distinguish real from simulated data, thus ensuring that they are treated equally.  The DAQ data is held in an object known as TrackerDAQ, within which the data is sub-divided according to the DAQ system it originated from.

The Monte Carlo event holds data on ``scifi hits'' produced by tracks passing through the fibre planes and any noise hits originating in those planes. The scifi hit is implemented as a class based on the generic hit-class template from which all the different MICE detector-hit classes are derived.  Other relevant data held in the Monte Carlo event, though not part of the tracker data structure, include the simulated tracks and ``virtual hits'' which contain information of the track state at their location. Such data is used to evaluate the reconstruction performance against the original simulated data (see section~\ref{sec:Performance}).

\subsection{Tracker reconstruction data structure}
\label{subsec:TrackerReconDataStructure}
The reconstructed data for the tracker is held in the ``scifi event'' class.  This contains vectors of the following container classes that represent the higher-level reconstructed tracker-data:

\begin{itemize}
  \item ``Digits'' contain the analogue-to-digital converter (ADC) counts (real or simulated) from the readout of a single channel in response to an incident track;
  \item ``Clusters'' are groups of neighbouring digits arising from a particle crossing one or two channels;
  \item ``Spacepoints'' group clusters from adjacent detector planes to give a point in tracker coordinates;
  \item ``Straight pattern recognition tracks'' group together spacepoints from different tracker stations when the track that produced the hits was straight (i.e. when the magnetic field is off); 
  \item ``Helical pattern recognition tracks'' group together spacepoints from different tracker stations when the track that produced the hits was helical (i.e. when the magnetic field is on); 
  \item ``Scifi tracks'' hold the final Kalman fit parameters of the particle track; and
  \item ``Trackpoints'' hold the fit parameters at each detector plane, including the momentum and position of the track.
\end{itemize}

Each of these objects is stored directly in the scifi event object, with the exception of trackpoints which are stored in the associated scifi tracks.  Each higher-level object also contains cross links in the form of pointers back to the objects within the scifi event which were used to create it. In this manner all higher-level objects can be traced back to the original digits.  In the case of a Monte Carlo run the digits themselves are linked via an ID number and lookup table back to the scifi hits used to produce them. The ID is defined as \textit{tspc}, where \textit{t} is the tracker number, \textit{s} is the station number, \textit{p} is the plane number and \textit{c} is the channel number (given with three numerals e.g. 010 for channel 10).  This structure means the reconstruction branch has no direct reference to the Monte Carlo data.
\section{Geometry}
\label{sec:Geometry}
  
  \begin{table} [tbp]
  \begin{center}
  \begin{tabular} {|c|c|c|c|c|c|}
    \hline
    \multicolumn{6}{|l|}{Tracker 1 offsets in mm} \\
    \hline
    & Station 1 & Station 2 & Station 3 & Station 4 & Station 5 \\
    \hline
    X & 0.0 & -0.5709 & -1.2021 & -0.5694 & 0.0 \\
    Y & 0.0 & -0.7375 & -0.1657 & -0.6040 & 0.0 \\
    Z & -1099.7578 & -899.7932 & -649.9302 & -349.9298 & 0.0 \\
    \hline
    \hline
    \multicolumn{6}{|l|}{Tracker 2 offsets in mm} \\
    \hline
    & Station 1 & Station 2 & Station 3 & Station 4 & Station 5 \\
    \hline
    X & 0.0 & -0.4698 & -0.6717 & 0.1722 & 0.0 \\
    Y & 0.0 & 0.0052 & -0.1759 & -0.2912 & 0.0 \\
    Z & -1099.9026 & -899.009 & -650.0036 & -350.0742 & 0.0 \\
    \hline
  \end{tabular}
  \caption{\label{tab:CMM} The position of the tracker stations with respect to the tracker reference surface as measured by the coordinate measuring machine.}
  \end{center}
  \end{table}
  
  The position of each tracker station was determined with respect to the tracker reference surface using a coordinate measuring machine (see table~\ref{tab:CMM}). The station positions are stored in the MICE configuration database (CDB). The CDB is a bi-temporal database, alterations being tracked by date and run number~\cite{DavidForrestThesis}. Information is stored in the CDB as a collection of XML files which are translated into the native MAUS format ``MiceModules'', at run-time.  The MiceModules are text documents and contain all the information needed to simulate the various MICE systems and detectors.  MAUS uses the same geometry descriptions for both simulation and reconstruction. 
  
  In the description of the geometry MAUS adopts a passive rotation convention to be consistent with GEANT4.  The active volume of each tracker is given by a cylinder of 150~mm radius, which is used to define the fiducial volume for the reconstruction. Alignment of the individual tracking stations and the trackers themselves to the solenoid axis has been completed using real-data.
  

\section{Simulation}
\label{sec:Simulation}

The simulation of the trackers makes use of the GEANT4 standard physics libraries to describe particle motion through the fields and material of the beamline. The trackers are simulated on a per-fibre basis and arranged into doublet-layer planes (as described in section~\ref{subsec:Trackers}). As particles pass through the fibres scifi hits are generated, containing the energy deposited in the fibre. 

The scifi hits are converted by the MAUS module used to digitise simulated tracker data (MapCppTrackerMCDigitisation) into to a number of photo-electrons (NPE) produced in the tracker VLPCs by means of a simple conversion factor. At this point the NPE values are non-integer, and are quantised simply by rounding down. These NPE values are then ``smeared'' to simulate the detector response (the electron avalanche effect in the VLPCs). The smearing process involves modeling this response as a Gaussian, the mean being given by the quantised NPE value and the width being determined from data. Once the smearing is complete the signal is split into $2^8$ bins to represent the sampling of the 8 bit ADCs. The simulated ADC counts are then used together with the measured calibration of each channel to give a final NPE value.

It is also possible to add noise to the digitisation process through the addition of an extra MAUS module (MapCppTrackerMCNoise). Noise arises from thermally excited electrons in the VLPCs. The process is stochastic and the rate is controlled by altering the bias voltage on the VLPCs to give 1 NPE in 1.5\% of fibres per particle trigger. The noise is generated as a Poisson distribution. The effect is modelled in the software and used to introduce additional photo-electrons to the simulated signal prior to the quantisation and smearing stage. 

Once the final NPE value has been calculated it is combined with the channel number to form a digit object. The digits are then added to the scifi event and sent on to the reconstruction modules.

\section{Reconstruction}
\label{sec:Reconstruction}
\begin{figure}[tb]
  \begin{center}
    \includegraphics[width=0.95\linewidth]{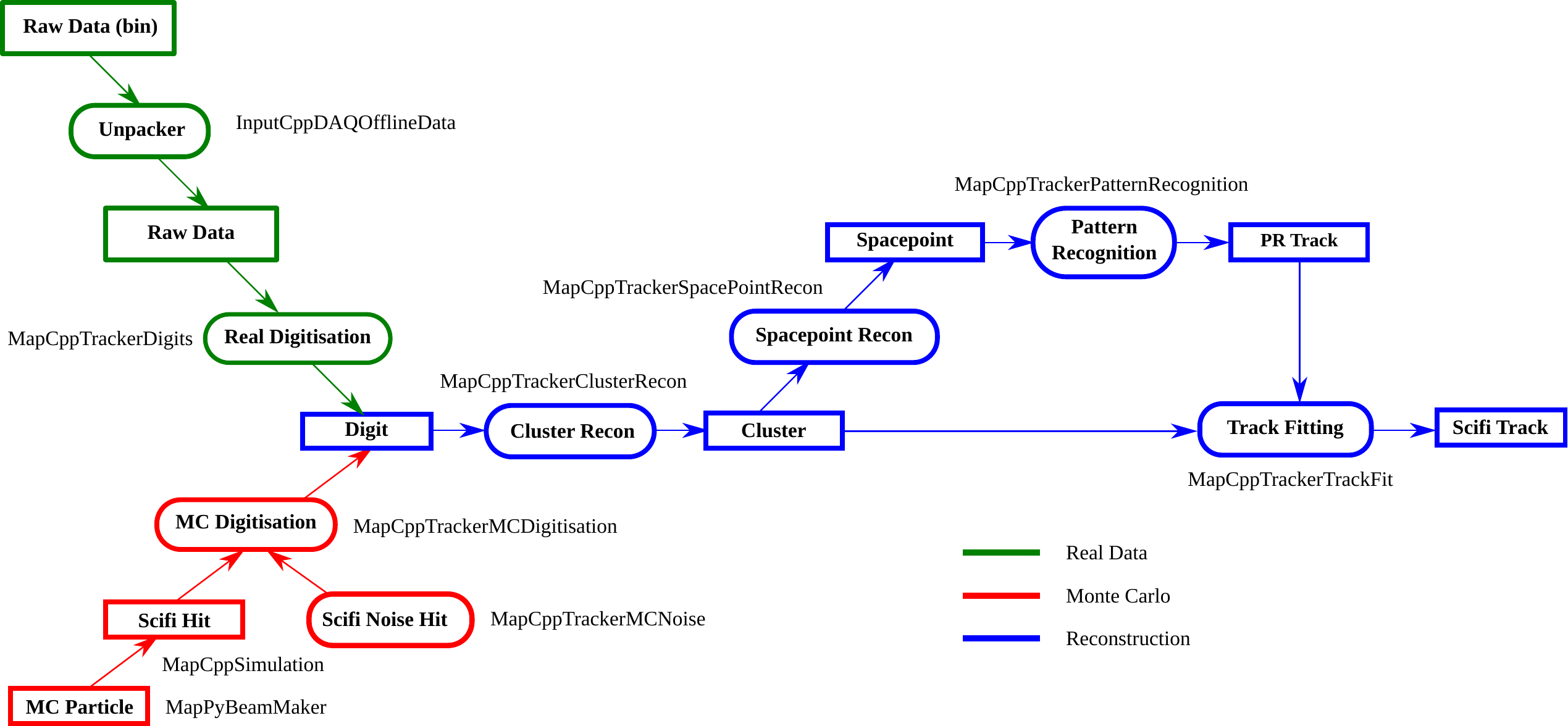}
    \caption{\label{fig:DataFlow} The reconstruction data flow. Data originates either from simulated or real-data, the two branches merge after digitisation, after which the reconstruction proceeds identically.  The relevant MAUS modules for each step are indicated.}
  \end{center}
\end{figure}

The reconstruction process begins with either real or simulated data and proceeds to reconstruct progressively higher-level objects step-by-step, culminating in scifi tracks and trackpoints. The digits, real or simulated, are passed to the same set of MAUS modules and the reconstruction proceeds identically for either case. The reconstruction process is illustrated in figure~\ref{fig:DataFlow} and is described in the sections that follow.

  \subsection{Digitization}
  \label{subsec:Digitization}
  For real data the signals from the tracker ADCs are recorded by the DAQ system.  Channel-by-channel calibration constants are used to convert the ADC value to a signal in NPE and the DAQ channel number to tracker channel number.  This information is then used to form a digit.  The analogous process for Monte Carlo data is described in section~\ref{sec:Simulation}.

  \subsection{Cluster Reconstruction}
  \label{subsec:Clustering}
  The clustering algorithm loops over every combination of pairs of digits in a scifi event and combines any that occur in neighbouring channels. In the case of a multi-digit cluster, the unweighted average channel value is used to define the plane coordinate, $\alpha$, and the NPE is summed.

  \subsection{Spacepoint Reconstruction}
  \label{subsec:SpacepointReconstruction}
  For each station the constituent planes are searched for clusters that can be used to form a spacepoint. Spacepoints are constructed from clusters from all three planes (a triplet spacepoint) or from any two out of the three planes (a doublet spacepoint). 

  To determine which clusters from each plane originate from the same track, ``Kuno's conjecture'' is used: for a given triplet spacepoint the sum of the channel numbers of each cluster will be a constant.  So if $n^u$, $n^v$ and $n^w$ are the fibre numbers of the clusters in $u$, $v$ and $d$ and $n^u_0$, $n^v_0$ and $n^w_0$ are the corresponding central channel numbers. Three clusters form a space point if:
  \begin{equation}
    | (n^u + n^v + n^w) - (n^u_0 + n^v_0 + n^w_0) | < K \, ;
  \end{equation}
  where $K$ is a constant, taken by default to be 3.0. The clusters are sorted according to NPE, with higher NPE clusters being matched using Kuno's conjecture first. 
  
  Once all triplet spacepoints have been found, doublet spacepoints are created from pairs of the clusters that remain, the only selection criterion applied being that the crossing point of the two channels is within the tracker radius. Spacepoint position is determined from the cluster measurements, $\alpha$ (see section~\ref{subsec:PlaneAndClusters}) and the known plane orientations for those measurements. 

%


  \subsection{Pattern Recognition}
  \label{subsec:PatternRecognition}

  Pattern recognition is based on looping over combinations of spacepoints from different stations and performing a fit using a linear-least-squares technique. There are separate algorithms for the straight-track (no field) case and the helical-track case.

   \subsubsection{Helical Pattern Recognition}
   \label{subsubsec:HelicalPatternRecognition}

   
   The helical pattern recognition is performed in cylindrical co-ordinates $(r, \phi, z)$, where $r$ is the helix radius, $\phi$ the turning angle in the transverse plane, and $z$ is the longitudinal coordinate. An additional coordinate, $s$, the distance the particle travels in the transverse plane, is also defined. The helix is then parameterised by: the circle centre $x_c$, $y_c$ and the radius, $r$, in the transverse plane; $s_0$ the value of $s$ where the helix crosses the tracker reference plane; and $t_s = ds/dz$, which describes the tightness of the coiling. 

   To find a track, one spacepoint is selected from each station and a circle is fitted in the $(r, \phi)$ projection. If the $\chi^2$ of this fit is sufficiently small (by default less than 15.0 multiplied by the number of degrees of freedom, NDF) then the value of $\phi$ is used to generate $s$. A straight-line fit is then performed in the $(z,s)$ plane and, if the $\chi^2$ in this projection is also small (by default less than 4.0 multiplied by the NDF), the track is accepted as a track candidate. Once all the different possible combinations of spacepoints have been tried the track candidate with the lowest combined $\chi^2$ is selected. Once every possible track with a spacepoint in all 5 stations has been found, the procedure is repeated looking for tracks with spacepoints in only 4 of the 5 stations. 
   

    \subsubsection{Straight Line Pattern Recognition}
    \label{subsubsec:StraightLinePatternRecognition}


    The straight-track finding is performed in Cartesian coordinates. The track parameters are $(x_0, y_0, t_x, t_y)$, where: $(x_0$, $y_0)$ is the position at which the track crosses the tracker reference surface; $t_x = dx/dz$; and $t_y = dy/dz$. Two spacepoints are chosen in the outer stations and a road is created between them. Any spacepoints in the road are fitted in the $(x,z)$ and $(y,z)$ planes. If the $\chi^2$ of the fits is small (by default less than 4.0 multiplied by the NDF) a candidate track is formed. Once all possible spacepoint combinations have been tried the track candidate with the lowest $\chi^2$ is selected. As in the helical case, following the completion of the full 5 point track search, tracks with spacepoints in 4 out of the 5 stations are searched for. For the straight case only, following the completion of the search for 4 point tracks, a search is also made for tracks with spacepoints in only 3 out of the 5 stations. 

   \subsection{Track Fit}
   \label{subsec:FinalTrackFit}
   The final track fit was implemented using a track-orientated Kalman filter~\cite{Fruhwirth,Billoir}, which can be shown to be an optimal linear fitter, that takes into account all correlations and measurements, for a linear system. The Kalman filter is an iterative algorithm that incrementally propagates an estimate of the current track state between measurement planes, using measurement information to ``filter'' the state, improving the estimate of the track state.
   
   For the helical-track fit, the system is only approximately linear, hence an extended Kalman filter~\cite{Goodwin} was implemented which analytically propagates the track states between measurements, while the covariance matrices are propagated using a first-order linear approximation to the non-linear system.  In contrast to pattern recognition, the Kalman Filter uses a different state-space model, ($x$, $p_x$, $y$, $p_y$, $q/p_z$) rather than ($x_c$, $y_c$, $r$, $s_0$, $t_s$). This permits energy loss and multiple Coulomb scattering to be modelled linearly and implmentated using a simple set of algorithms.

    A MAUS-specific implementation of the Kalman filter has been developed. In order to make the implementation flexible and re-usable the core algorithm has no dependencies on phase-space dimensions or physical effects. 
    The principal components of the implementation are: propagation routines for both helical and straight-tracks; a measurement routine that transforms the track state-space into the measurement state-space; and approximations for the process and measurement noise.
    
    Pattern recognition provides a set of clusters that are associated with a track, together with the fitted parameters of the track. These parameters are used to provide the seed for the Kalman fit and the cluster information is used as the measurement data. The flexibility of the Kalman algorithm permits the effects of individual planes (multiple Coulomb scattering and energy loss) and the effect of the helium gas within the tracker to be accounted for between each measurement point.
    
    As the measurements correspond to individual clusters, the cluster parameter $\alpha$ (see section~\ref{subsec:PlaneAndClusters}) forms the one-dimensional measurement state. The measurement noise corresponds to the statistical spread of measurements within a single channel of width $d$ in the tracker readout. If the channel is modelled as a top-hat function, the variance of the function is $d^2/12$. Therefore the measurement noise was taken to be $d/\sqrt{12}$ for all clusters.

    The process noise was implemented as a combination of multiple Coulomb scattering and energy straggling. The energy loss is calculated using the Bethe formula~\cite{PDG}, for the most probable energy loss, and applied during the propagation stage. The noise term itself is calculated per increment as an RMS scattering angle using an implementation of the Highland formula~\cite{Highland}, in a fashion almost identical to the GEANT4 implementation:

    \begin{equation}
      \theta_{RMS} = \frac{13.6\ \textrm{MeV/c}}{\beta c p} Z \frac{x}{X_0}\left( 1 + 0.038 \ln\left(\frac{x}{X_0} \right)\right);
      \label{equ:highland_formula}
    \end{equation}
    where $\theta_{RMS}$ is the RMS scattering angle through a finite length of material, $\beta c$, $p$ and $Z$ are the velocity, momentum and charge of the particle in question, $x$ is the distance travelled and $X_0$ the radiation length of the material. Although the RMS scattering angle does not form a Gaussian distribution, the first order approximation is believed to be sufficient.




\section{Performance}
\label{sec:Performance}

  A Monte Carlo simulation was used to determine the efficiency and resolution of the reconstruction algorithms. It was necessarily Monte Carlo based in order to compare the reconstructed events to the simulated ``truth'', thereby highlighting any inefficiencies and inaccuracies.

  The fitted transverse positions, $(x,y)$, and momenta, $(p_x, p_y, p_z)$, were compared to the Monte Carlo truth on an event-by-event basis. The reconstruction of the Monte Carlo data followed the same requirements that are used for the real-data reconstruction. The Monte Carlo truth data was stored at every tracker plane (via virtual hits) to permit a direct comparison with the reconstructed data. All comparisons were made at the tracker reference surface. 

  An artificial beam was generated with uniform distributions for both the longitudinal and transverse momenta. This ensured that the results were not biased by the incoming beam distribution and that the full  phase-space acceptance was probed with equal statistics. In order to remove unwanted ``non-physical'' particles, a cut was placed on the Monte Carlo truth data during the analysis phase to eliminate tracks with a large $p_t/p_z$ ratio. Tracks with a ratio greater than $0.5$ (that is, a $45^\circ$ angle with respect to the $z$ axis) were rejected from the analysis.

  \subsection{Kuno's Conjecture}
  \label{sec:performance:kunos_conjecture} 
\begin{figure}[p]
    \centering
    \includegraphics[width=0.85\textwidth, angle=0]{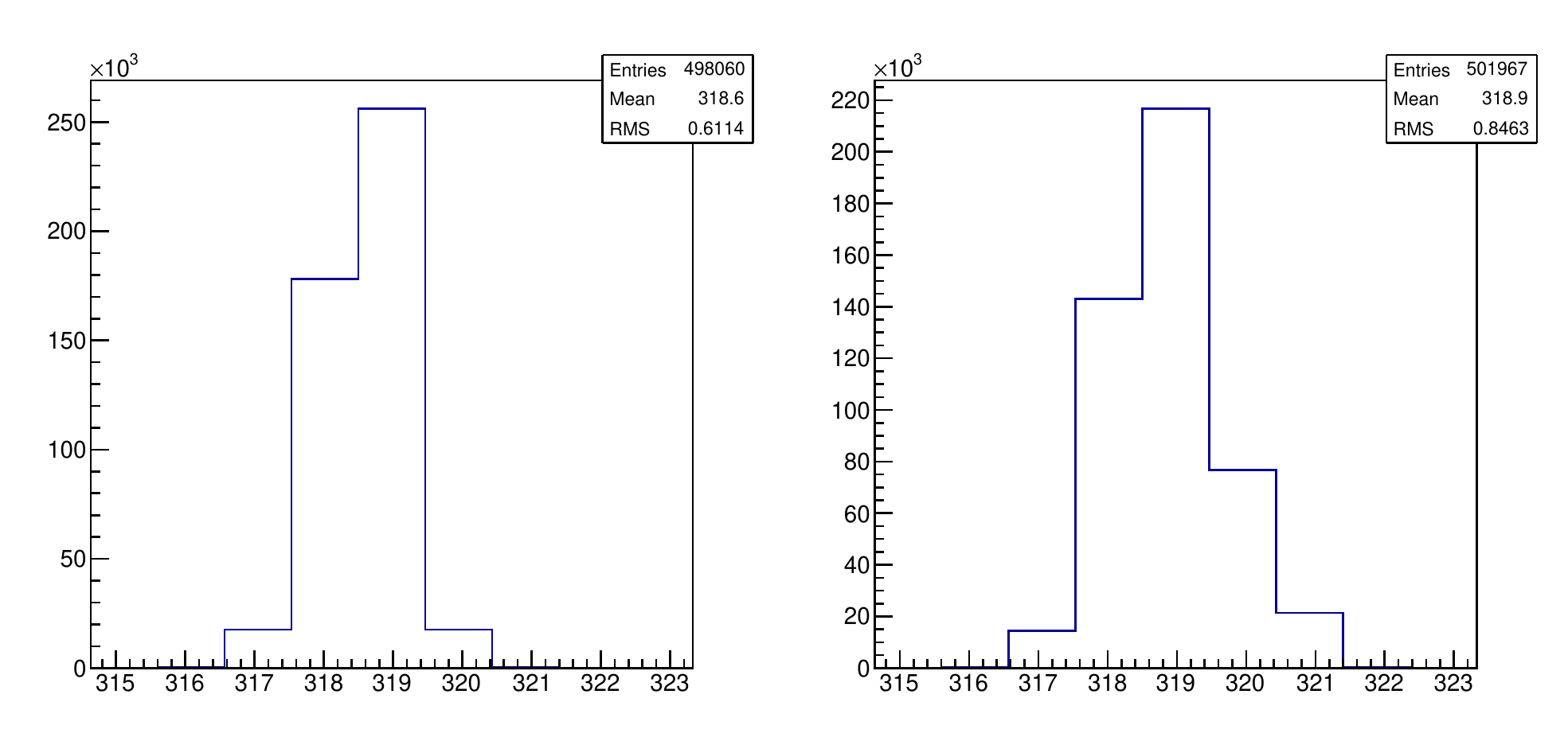}
    \caption{\label{fig:kuno} A plot showing the sum of the channel numbers for the clusters in each spacepoint for the upstream (left) and downstream (right) trackers. Kuno's conjecture that the sum is a constant is observed with a small variation in the downstream tracker arising from a slight fibre misalignment during construction.}
  \end{figure}

  All clusters selected to form triplet spacepoints should follow Kuno's conjecture (section~\ref{subsec:SpacepointReconstruction}). A plot showing the sum of the channel numbers for the clusters in each spacepoint is shown in figure~\ref{fig:kuno}. As expected the sum of the channel numbers for all the clusters in a spacepoint is constant to very good approximation. A distortion is visible in the downstream tracker, which arises from an offset introduced in the construction stage of one channel in one plane in one station. 
  

  \subsection{Track Finding Efficiency}
  \label{sec:performance:track_finding}
\begin{figure}[p]
    \centering
    \includegraphics[width=0.495\textwidth, angle=0]{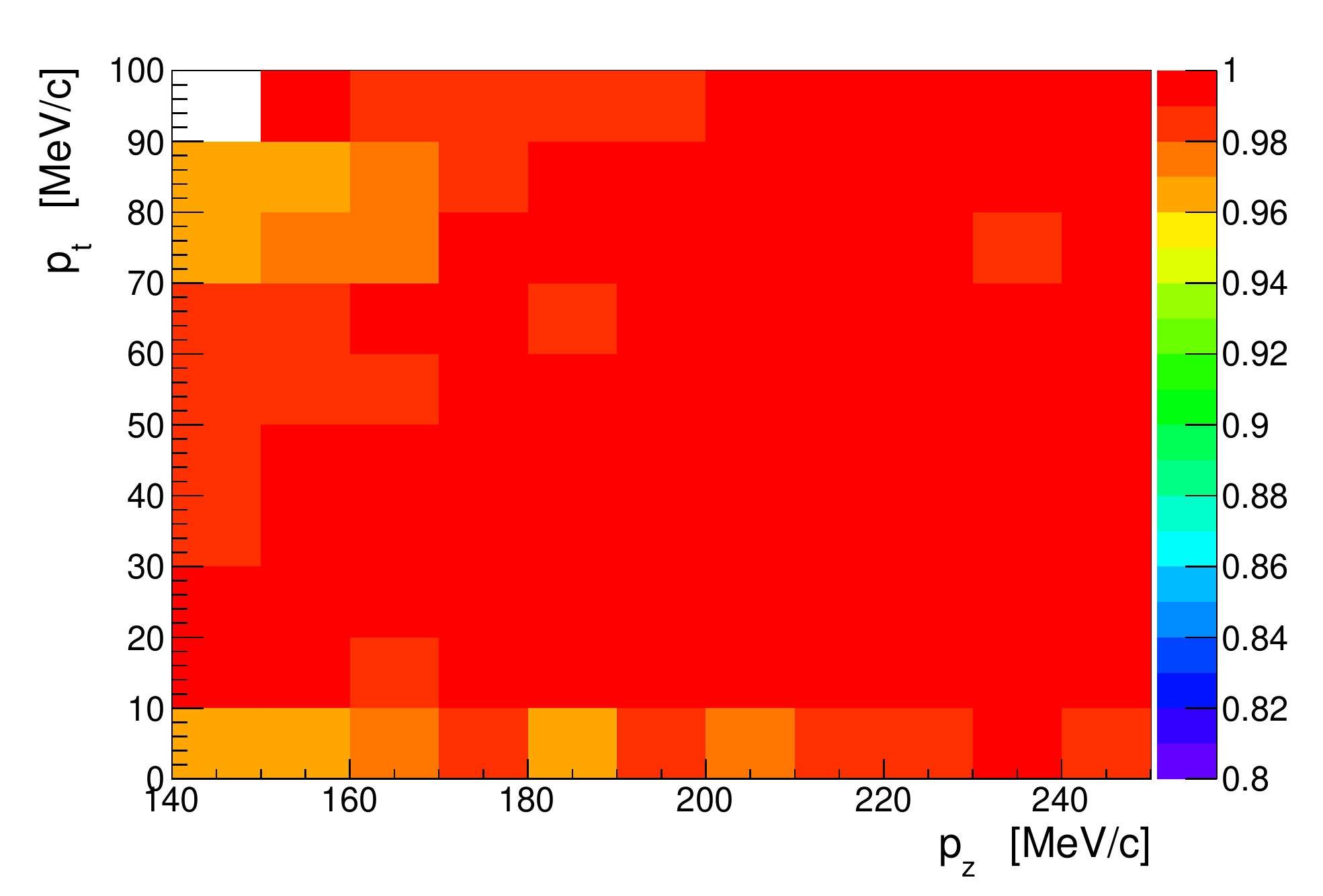}
    \includegraphics[width=0.495\textwidth, angle=0]{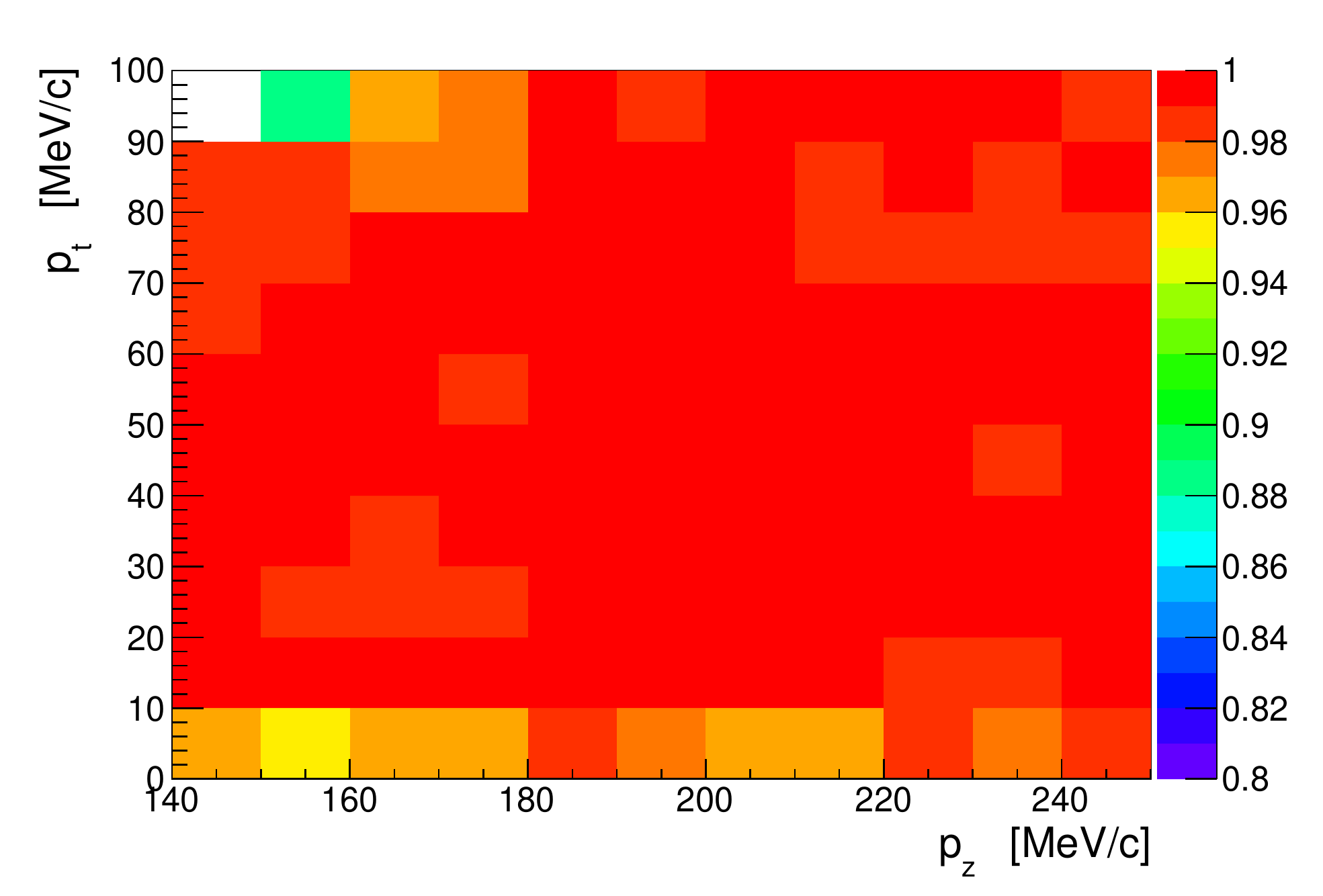}\\
    \caption{\label{fig:track_efficiency} The efficiency of reconstructing tracks in the upstream (left) and downstream (right) trackers as a function of the simulated longitudinal and transverse momentum. The white area in the top left of the plot indicates events which fall outside of the $p_t/p_z$ ratio cut.}
  \end{figure}

  For every simulated event, the number of tracks expected was calculated from the Monte Carlo truth. If a simulated track crossed enough tracker planes to create a sufficient number of spacepoints (3 for straight tracks and 4 for helical tracks), a reconstructed track was expected. The parameters of the reconstructed tracks were compared to the expected track parameters. The efficiency of track finding as a function of the true longitudinal and transverse momentum is shown in figure~\ref{fig:track_efficiency}. High efficiency is observed across the momentum range, with the exception of a dip at low transverse momentum ($p_t <$ 5~MeV/c) and a smaller dip in the high transverse momentum, low longitudinal momentum region. The former effect is an artifact of the track model at small track radii. Both effects are under investigation in order to improve the track finding efficiency in these regions.


  \subsection{Position and Momentum Resolution}
  \label{sec:performance:resolutions}
  
\begin{figure}[p]
   \centering
    \includegraphics[width=0.49\textwidth, angle=0]{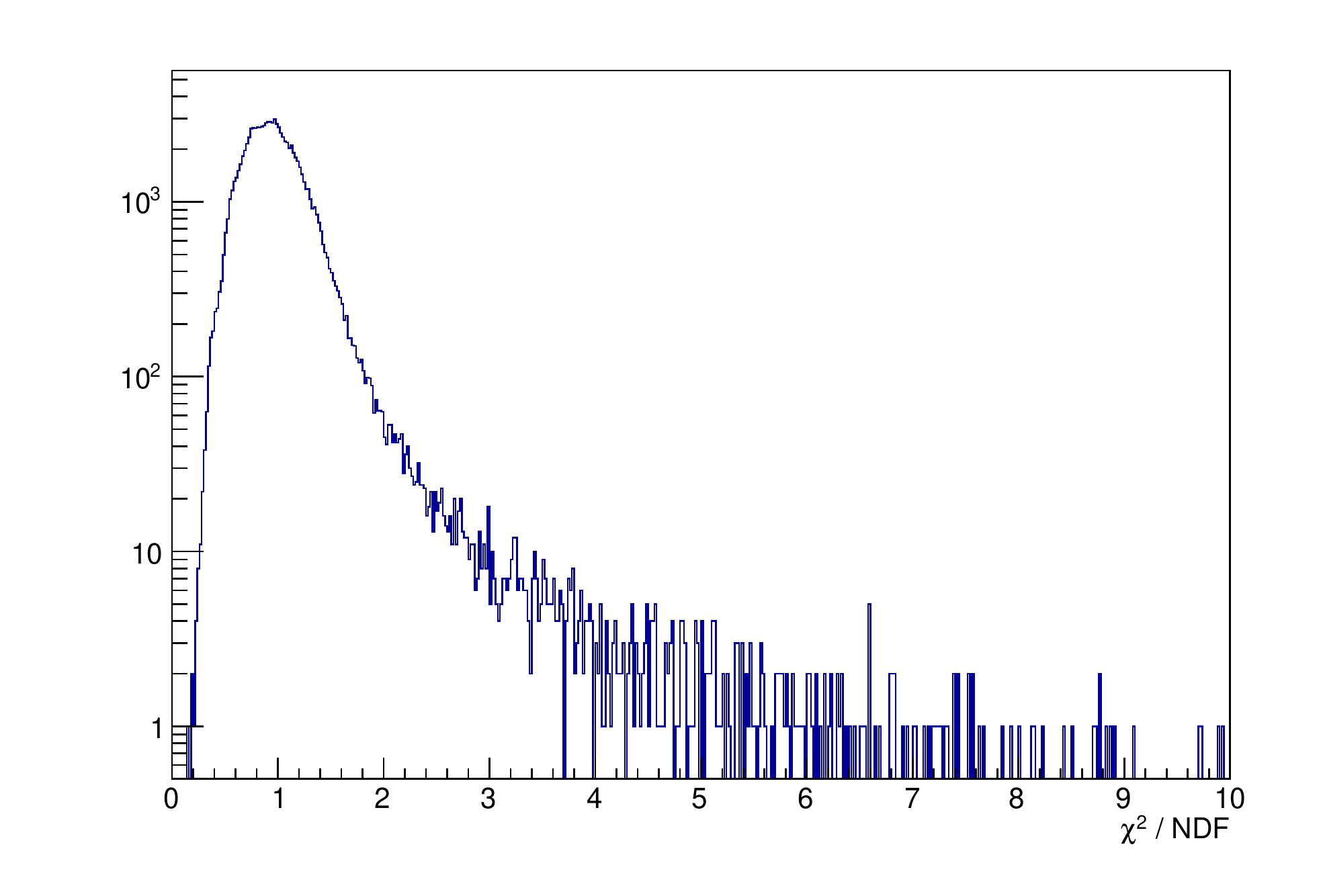}
    \includegraphics[width=0.49\textwidth, angle=0]{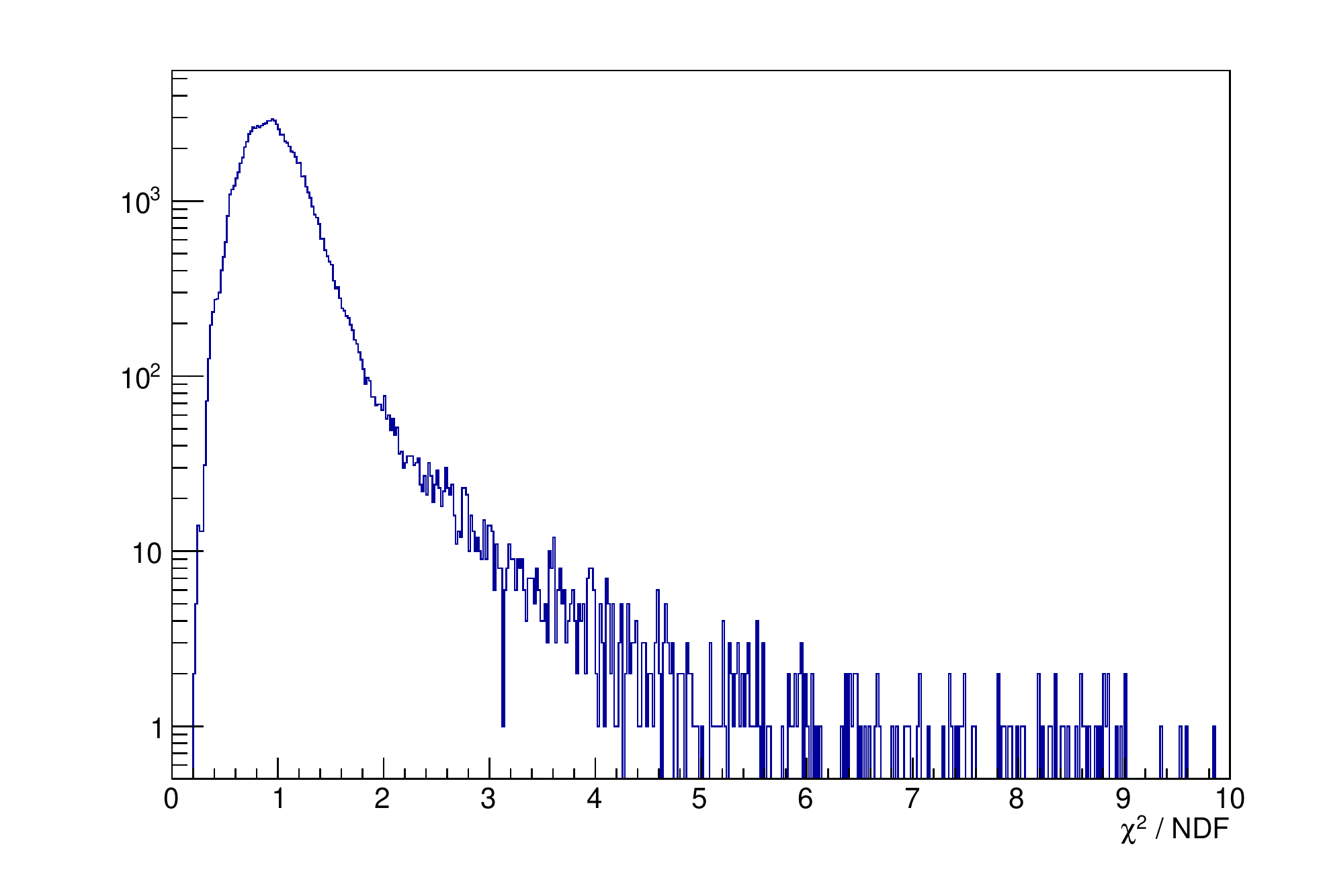}
   \caption{\label{fig:track_chisq} The $\chi^2$ per degree of freedom of the final track fit in the upstream (left) and downstream (right) trackers.}
  \end{figure}

\begin{figure}[p]
    \begin{center}
      \includegraphics[width=0.49\textwidth, angle=0]{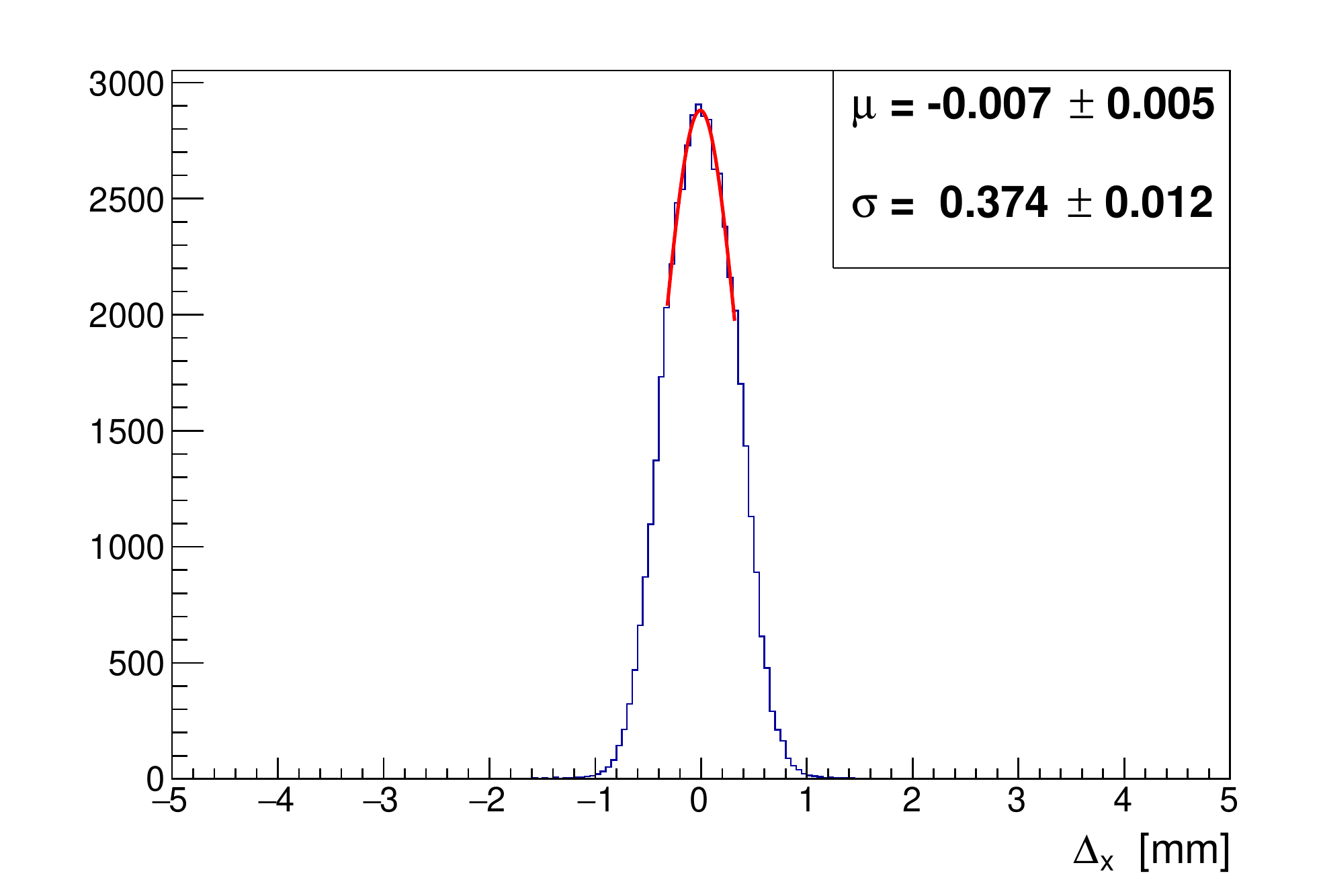}
      \includegraphics[width=0.49\textwidth, angle=0]{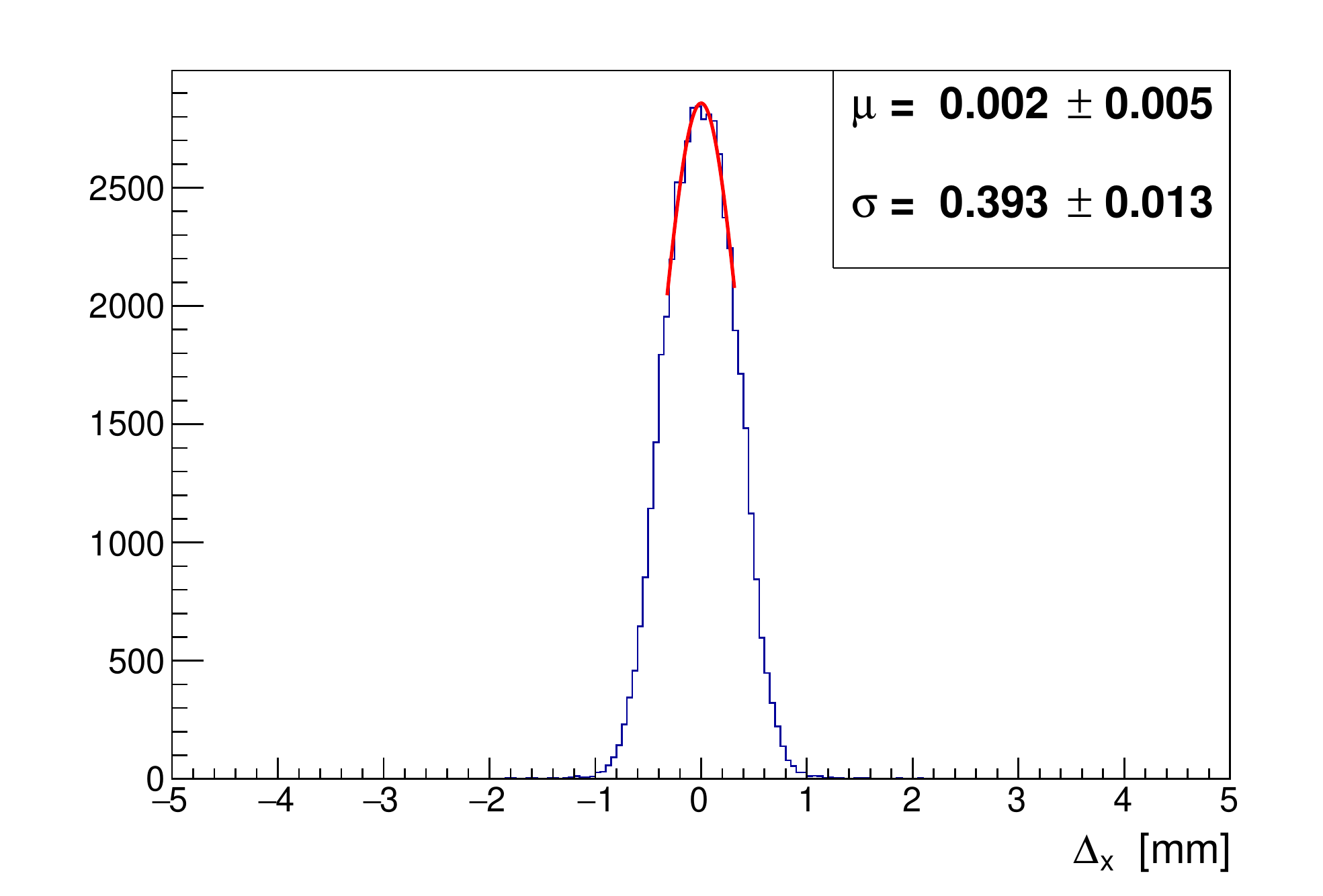}
      \caption{\label{fig:XResidKalman} The $x$ residuals of the upstream (left) and downstream (right) trackers.}
    \end{center}
  \end{figure}

\begin{figure}[p]
    \begin{center}
      \includegraphics[width=0.49\textwidth, angle=0]{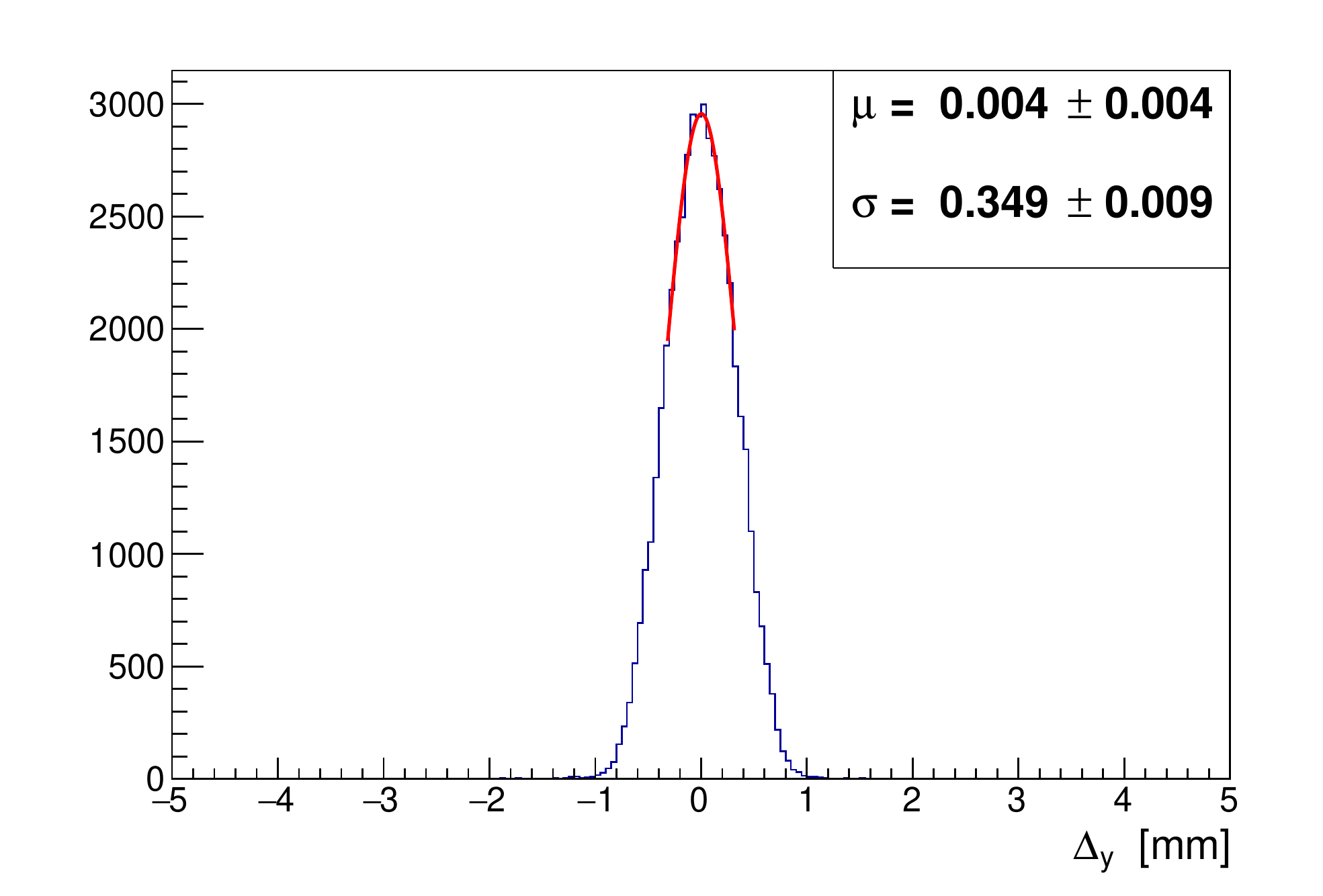}
      \includegraphics[width=0.49\textwidth, angle=0]{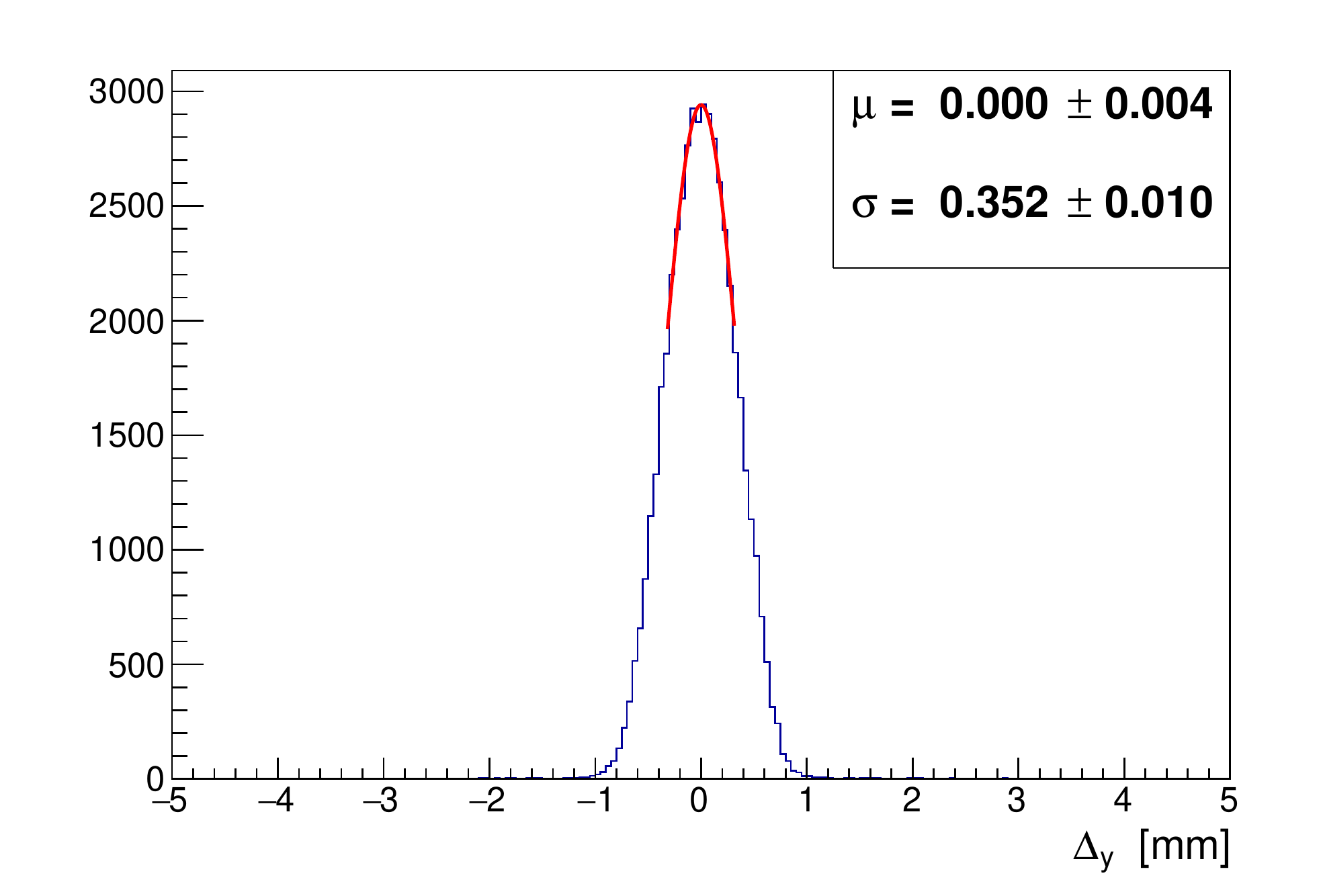}
      \caption{\label{fig:YResidKalman} The $y$ residuals of the upstream (left) and downstream (right) trackers.}
    \end{center}
  \end{figure}

\begin{figure}[p]
    \begin{center}
      \includegraphics[width=0.49\textwidth, angle=0]{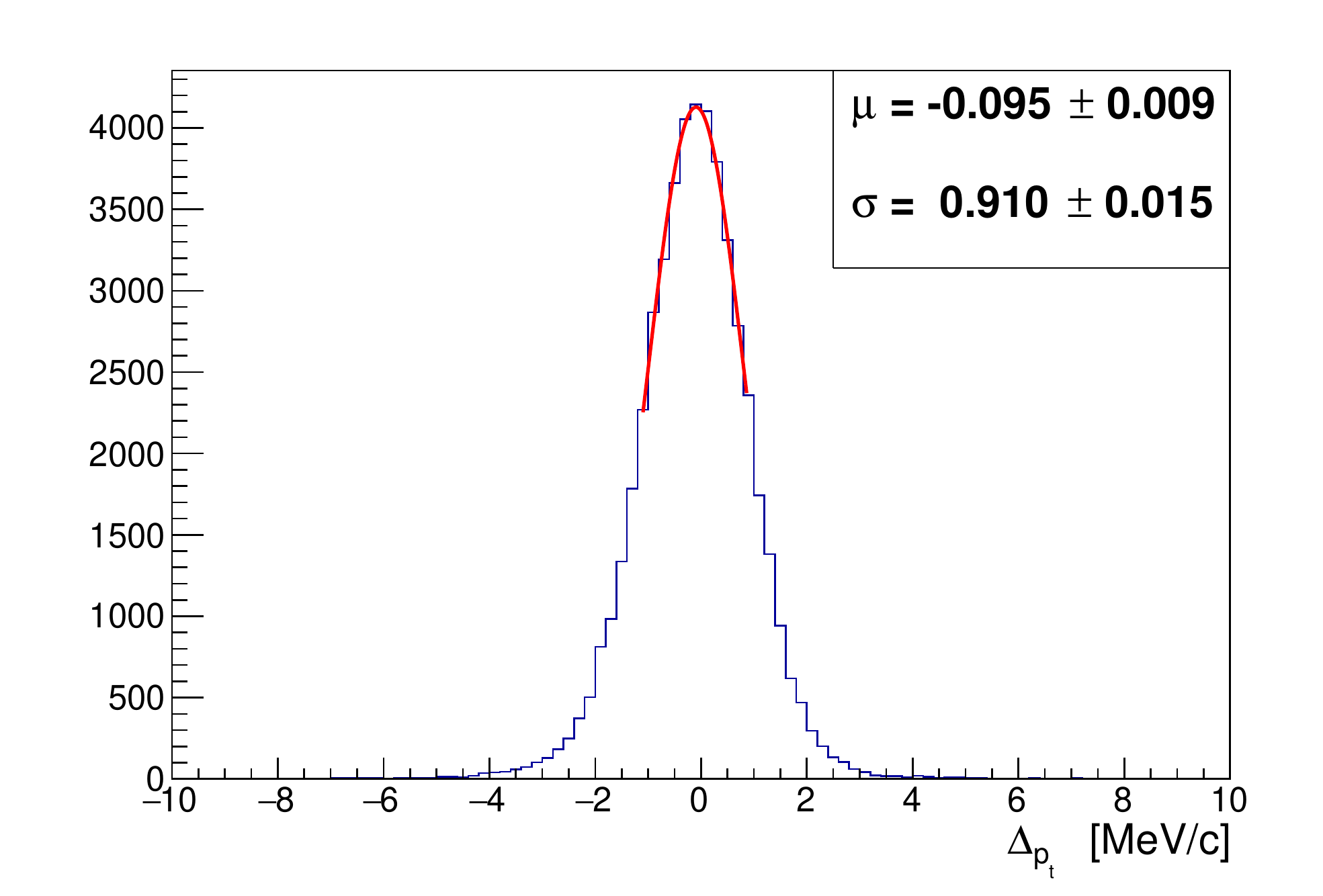}
      \includegraphics[width=0.49\textwidth, angle=0]{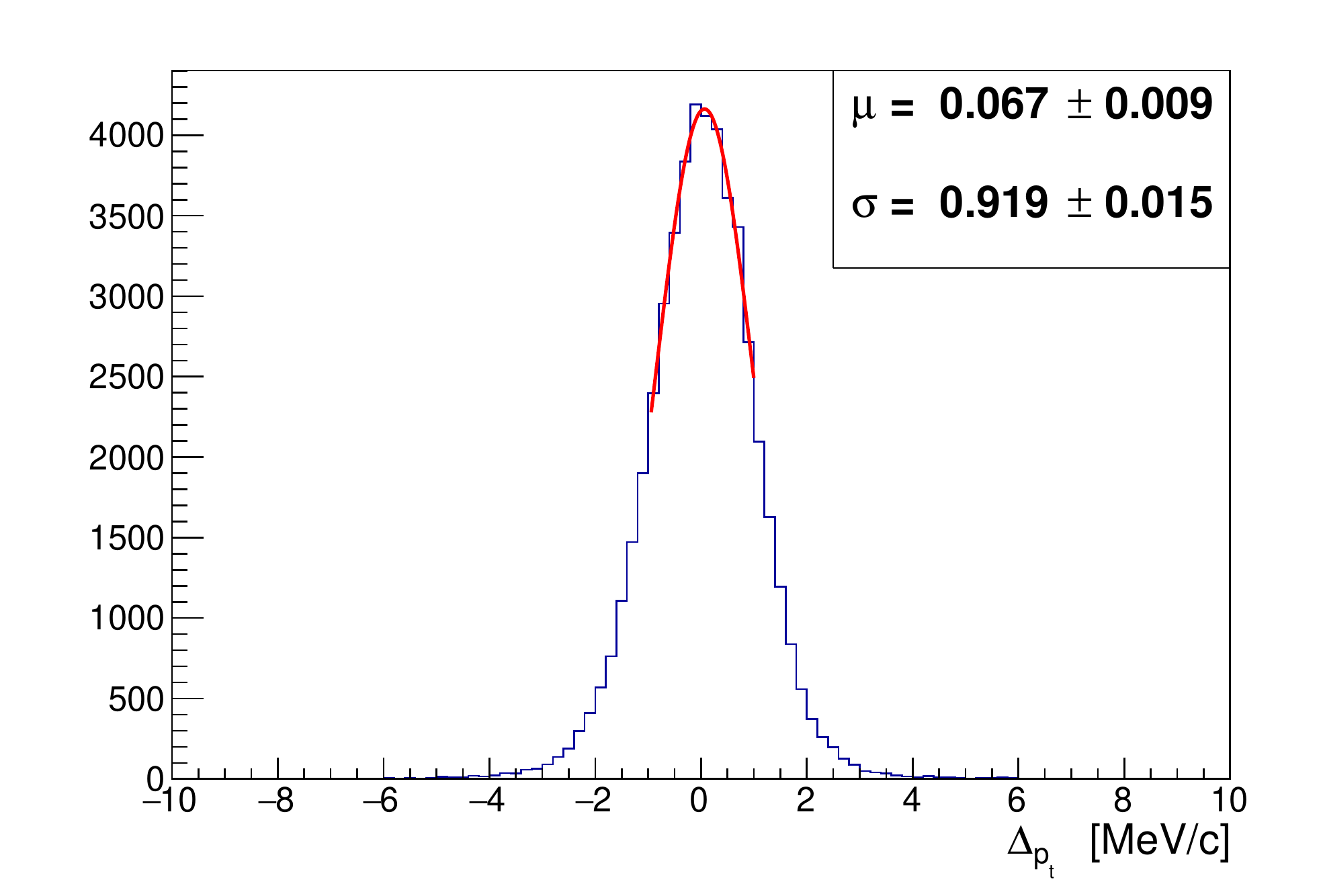}
      \caption{\label{fig:PtResidKalman} The $p_{t}$ residuals of the upstream (left) and downstream (right) trackers.}
    \end{center}
  \end{figure}

\begin{figure}[p]
    \begin{center}
      \includegraphics[width=0.49\textwidth, angle=0]{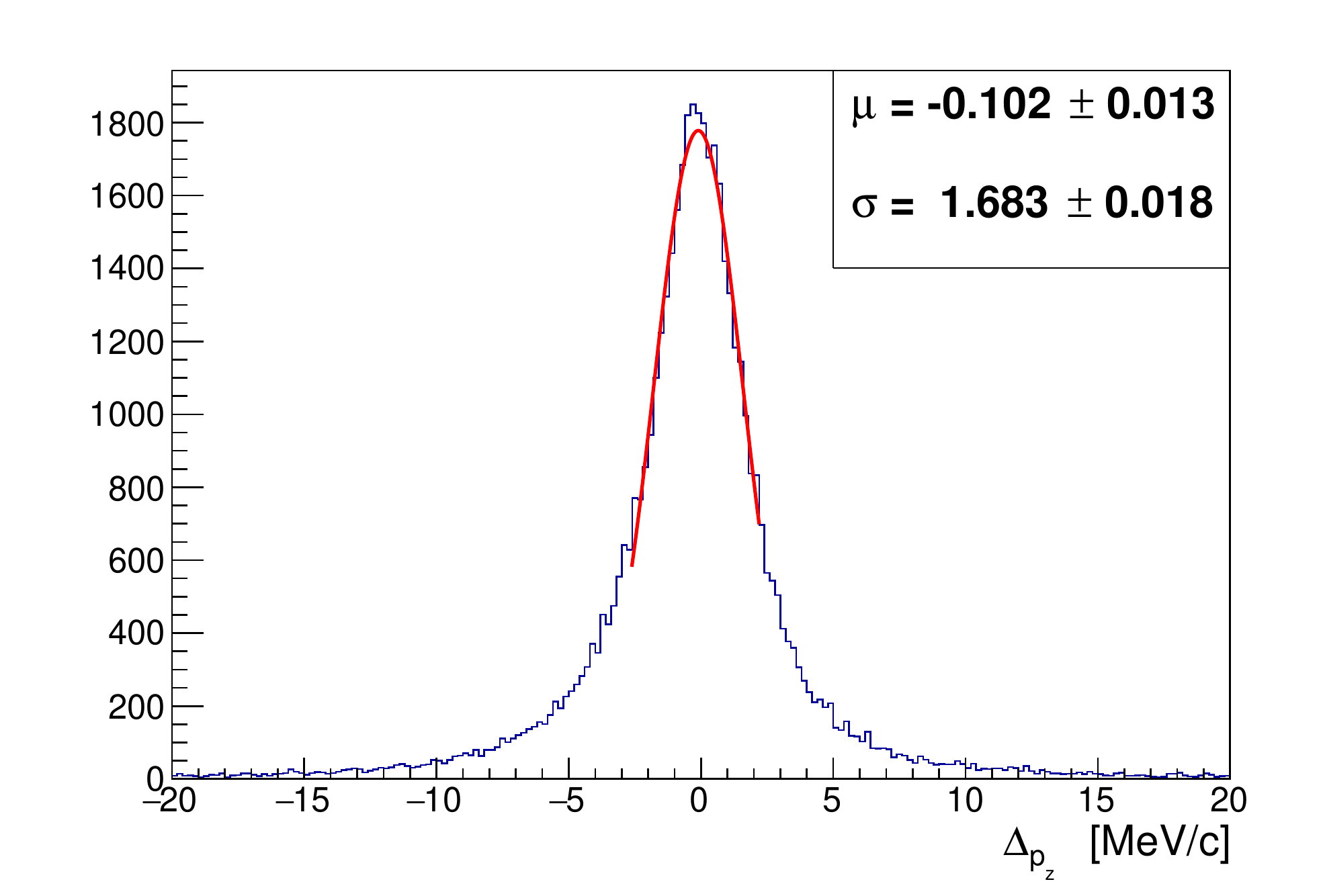}
      \includegraphics[width=0.49\textwidth, angle=0]{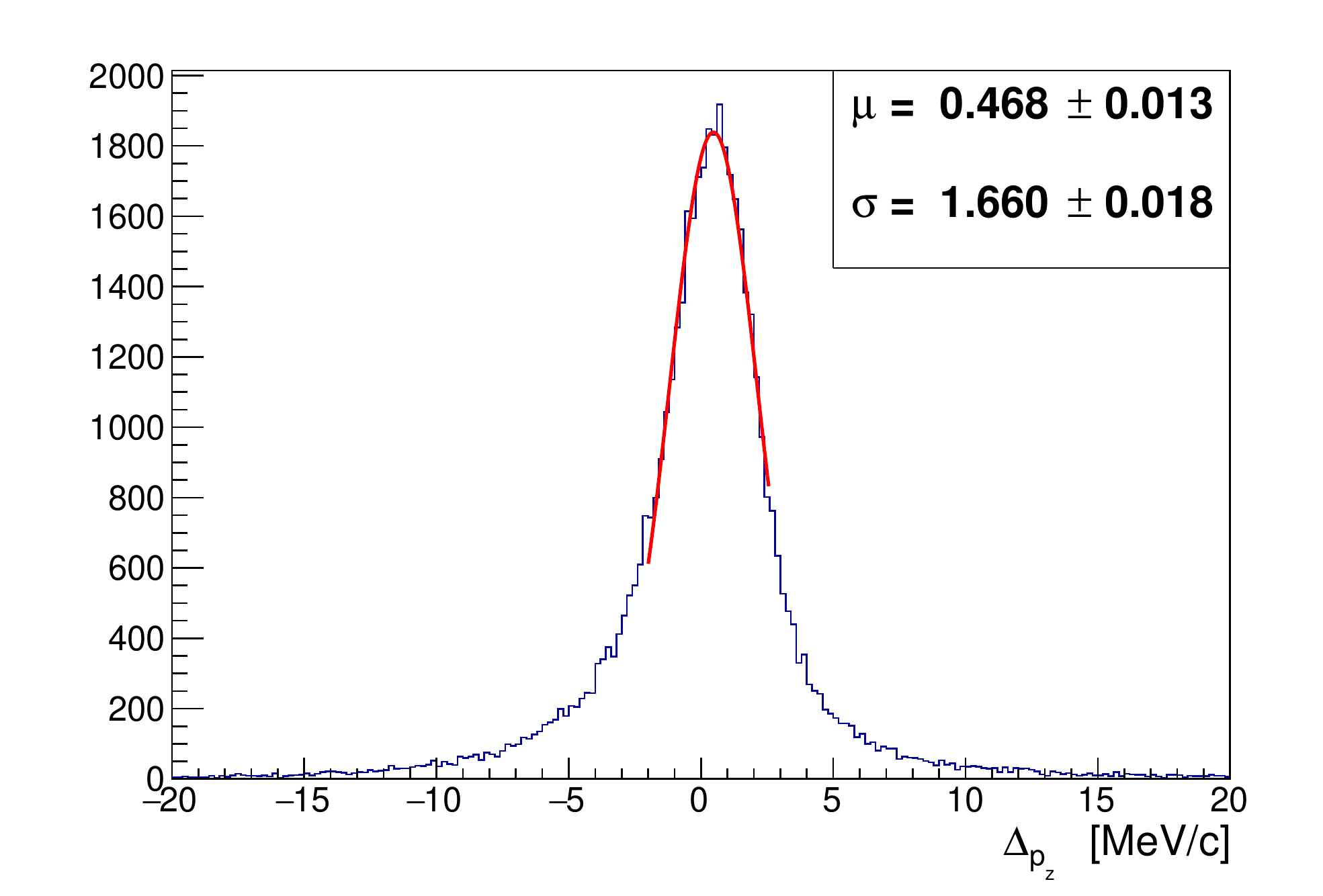}
      \caption{\label{fig:PzResidKalman} The $p_z$ residuals of the upstream (left) and downstream (right) trackers.}
    \end{center}
  \end{figure}

  
  The $\chi^2$ per degree of freedom for the track fits, used to provide the position and momentum determinations, are shown in figure~\ref{fig:track_chisq}. 
  The position residuals, shown in figures~\ref{fig:XResidKalman} and \ref{fig:YResidKalman}, are consistent with the expected measurement resolutions for a combined fit. The transverse momentum resolution, shown in figure~\ref{fig:PtResidKalman}, is consistent across the range of the sensitive phase-space at $\sim$0.9~MeV/c in both trackers. The longitudinal momentum shown in figure~\ref{fig:PzResidKalman}, an intrinsically more difficult measurement for the tracker, retains an acceptable spread of ${\sim4}$~MeV/c in both trackers. There is however a small systematic effect visible in the momentum distributions. This will be removed by matching the tracks to the instrumentation upstream and downstream of the tracker modules.

  In order to produce these plots, a requirement that there was a cluster within the reference plane was applied. Due to the effects of multiple Coulomb scattering, on rare occasions a single hard scatter can cause pattern recognition to miss a single spacepoint at the reference plane, hence creating a tail that will adversely affect the distributions. As we are concerned with the resolution following a successful pattern recognition stage, these events were removed.
  
 
  
  \begin{figure}[p]
   \begin{center}
     \includegraphics[width=0.49\textwidth, angle=0]{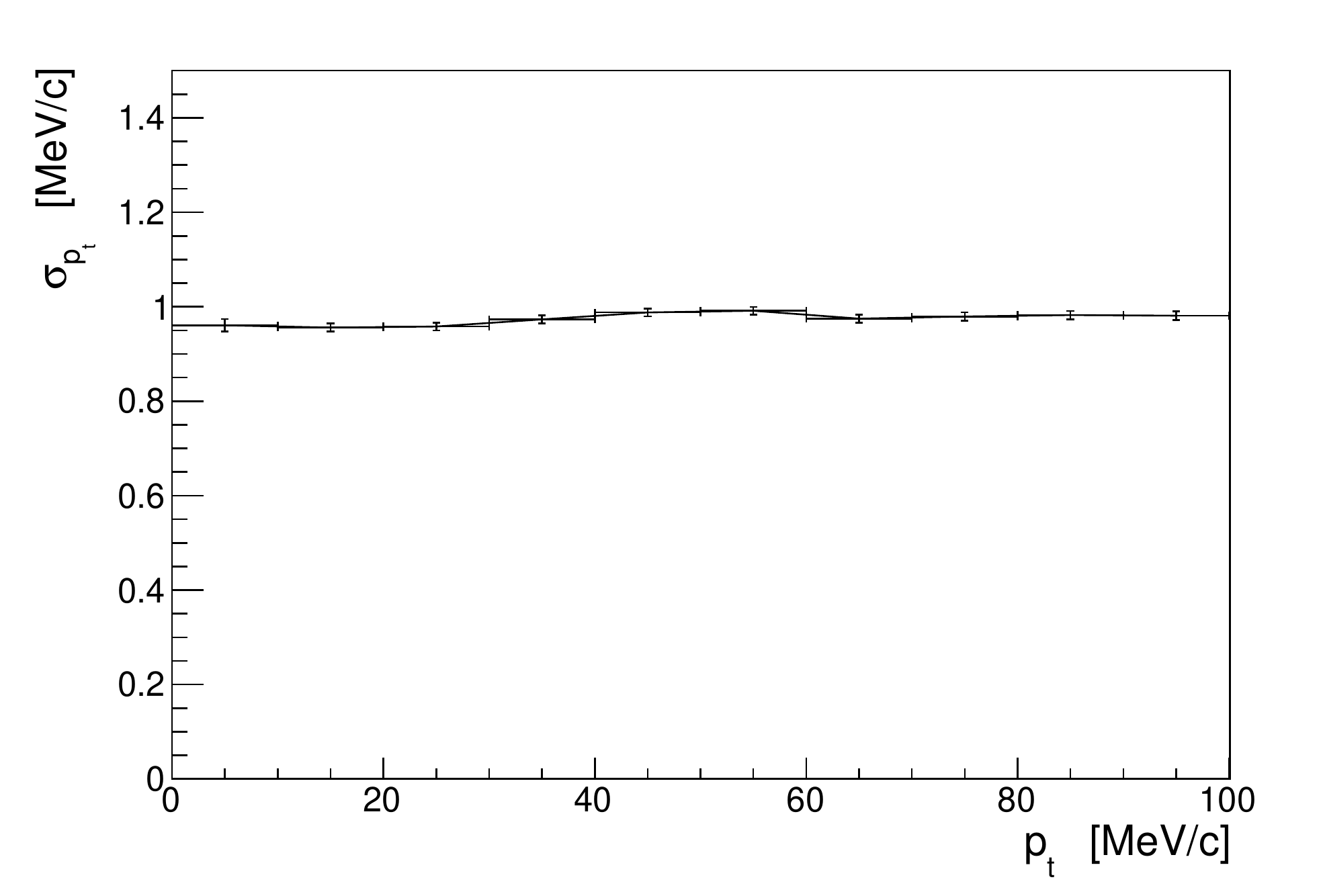}
     \includegraphics[width=0.49\textwidth, angle=0]{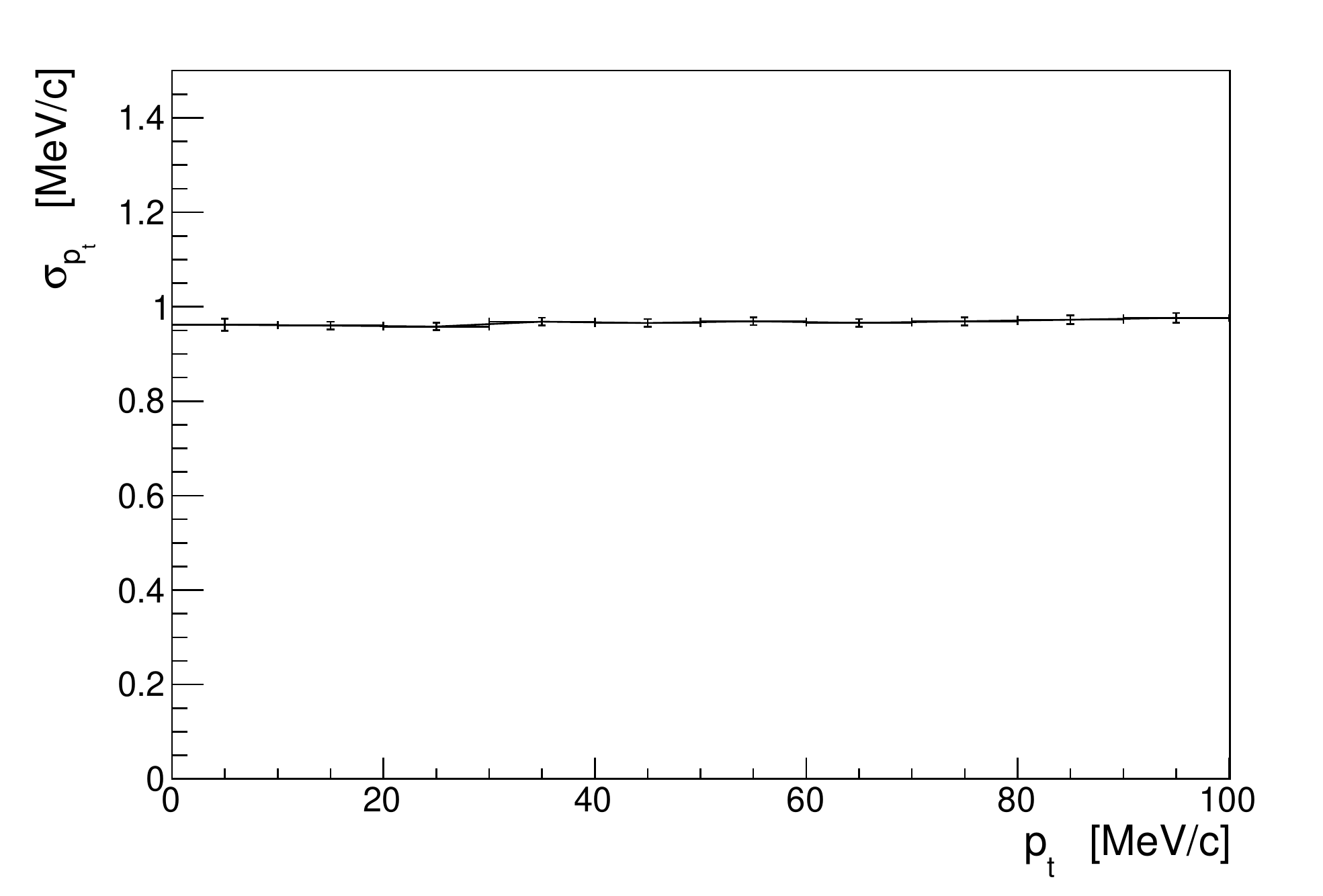}
     \caption{\label{fig:PtPtResolKalman} The $p_{t}$ resolution as a function of the $p_{t}$ of the upstream (left) and downstream (right) trackers.}
   \end{center}
  \end{figure}
  
  \begin{figure}[p]
   \begin{center}
     \includegraphics[width=0.49\textwidth, angle=0]{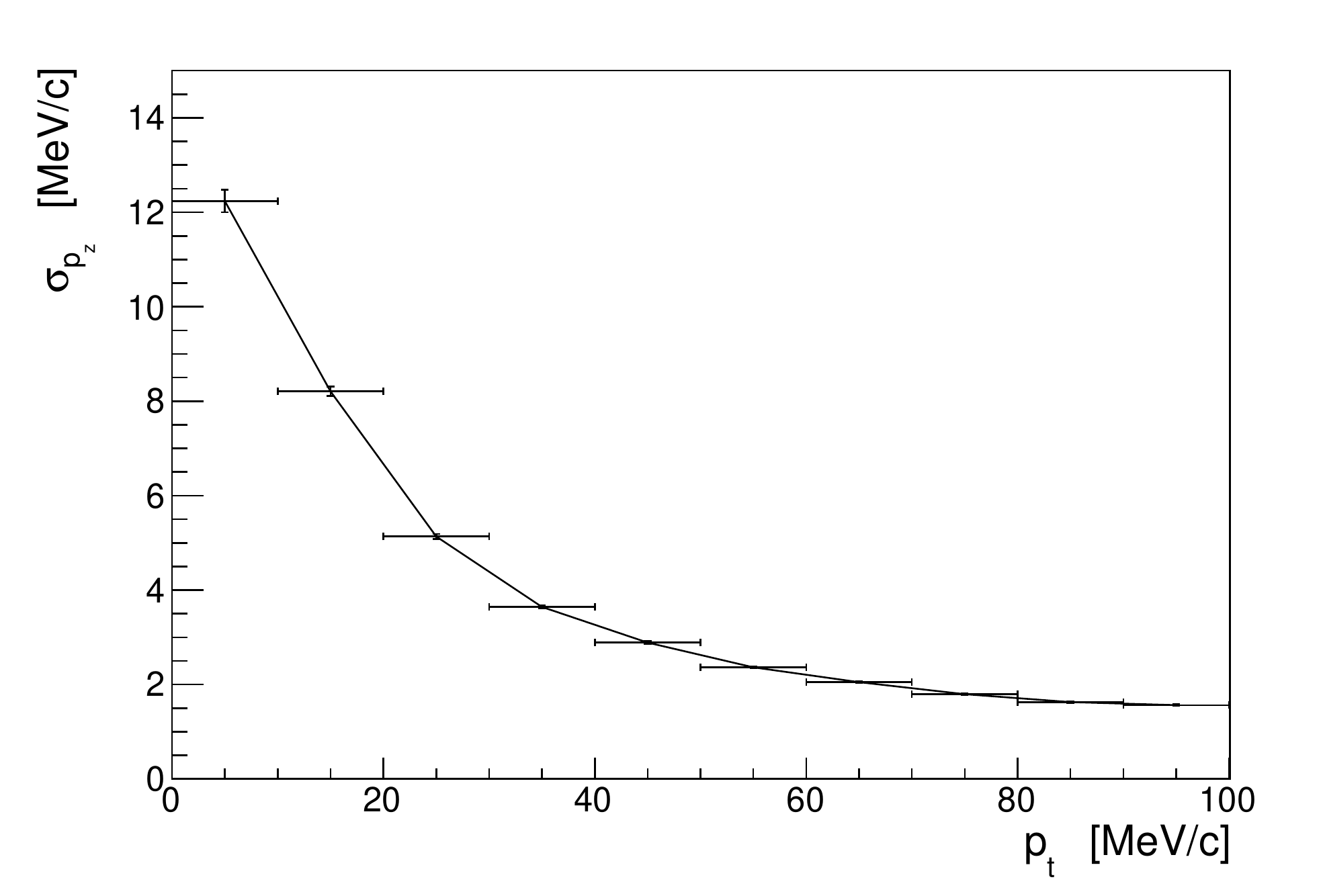}
     \includegraphics[width=0.49\textwidth, angle=0]{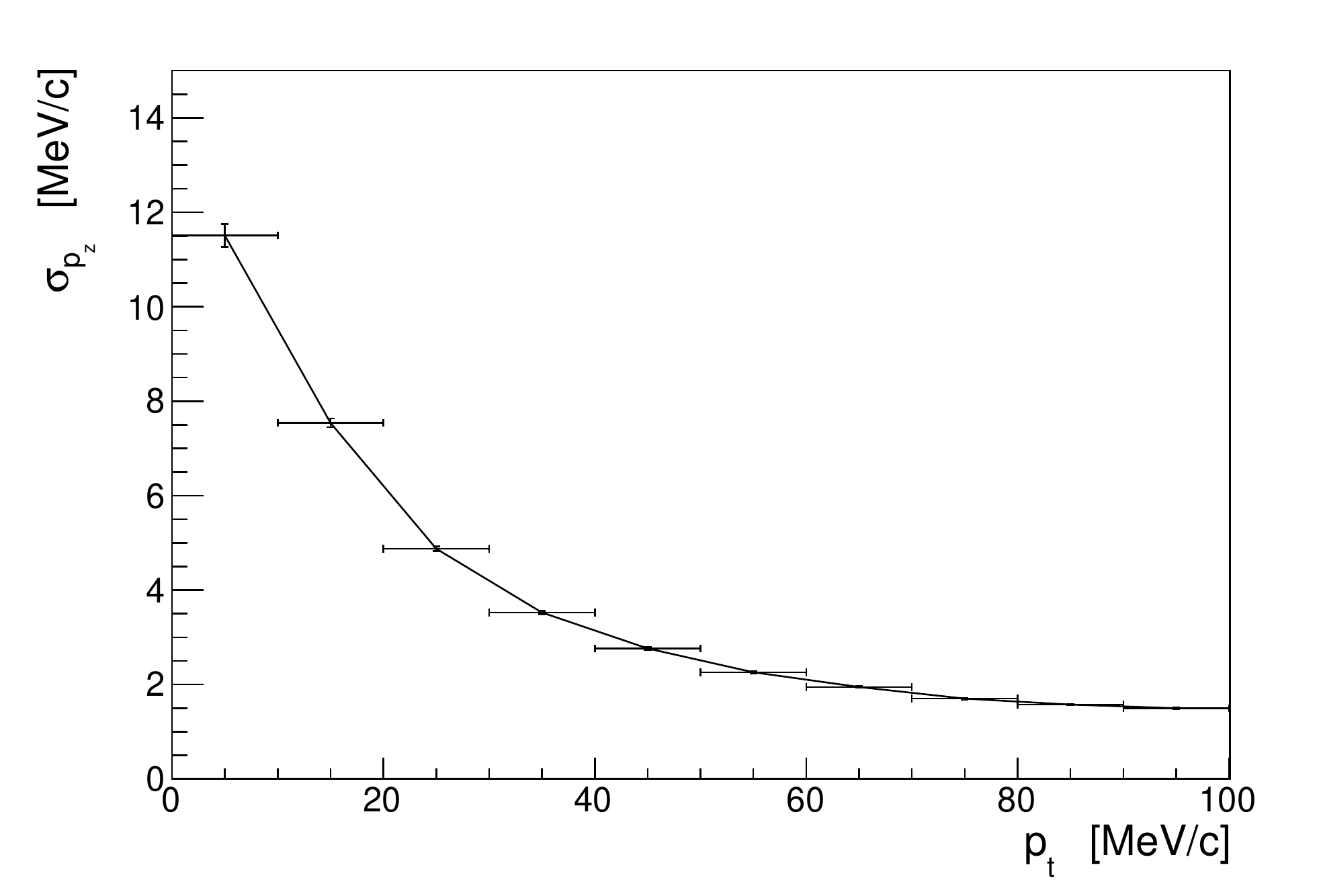}
     \caption{\label{fig:PtPzResolKalman} The $p_z$ resolution as a function of the $p_{t}$ of the upstream (left) and downstream (right) trackers.}
   \end{center}
  \end{figure}

\section{Conclusion}
\label{sec:Conclusion}
The tracker software has been presented and the performance of the reconstruction algorithms, including: spacepoint reconstruction; pattern recognition; and the final track fit, has been shown to meet expectation.

The performance of the final track fit was evaluated by comparing Monte Carlo truth with reconstructed data, for the key tracker measurements of $x$, $y$, $p_{t}$ and $p_z$.  The observed performance in the transverse position measurements is excellent, for both the upstream and downstream trackers. The reconstruction resolution in $p_t$ and $p_z$ meet specification and will provide a sufficient degree of precision for the MICE physics program. 

The track model uses a simplification of the energy loss in the tracker planes which leads to the systematic shift discussed in section 8.3. Since the two trackers were constructed to be identical this shift will cancel in the calculation of energy loss. Monte Carlo studies will allow us to calculate a linear correction to the measured momentum and account for the small remaining shift. We will use measurements by other detectors in MICE to validate this correction.

The tracker software is now used routinely in MICE data reconstruction, providing data for analysis and publication.

\section{Acknowledgements}
\label{sec:Acknowledgements}

The work described here was carried out in the context of the international Muon Ionization Cooling Experiment. It was made possible by grants from Department of Energy and National Science Foundation (USA) and the Science and Technology Facilities Council (UK). We gratefully acknowledge all sources of support. We also acknowledge the use of Grid computing resources deployed and operated by GridPP in the UK, http://www.gridpp.ac.uk/.


\bibliography{mice}                                                                           
\bibliographystyle{99-Styles/utphys}


\end{document}